\documentclass[fleqn,usenatbib]{mnras}

\usepackage{fix-cm}

\usepackage{lastpage}
\usepackage{newtxtext,newtxmath}
\usepackage{subcaption}

\usepackage{xcolor}

\usepackage[T1]{fontenc}

\DeclareRobustCommand{\VAN}[3]{#2}
\let\VANthebibliography\thebibliography
\def\thebibliography{\DeclareRobustCommand{\VAN}[3]{##3}\VANthebibliography}

\usepackage{graphicx}	
\usepackage{amsmath}

\title[Discovery of a radio flaring M dwarf]{Discovery of a radio-flaring M dwarf in a commensal transient search of the LADUMA field}

\author[M. Mlangeni et al.]{Moses Mlangeni$^{1,2}$\thanks{E-mail: m.mlangeni@blackgem.org},
Patrick~A.~Woudt$^{1}$,
Paul~J.~Groot$^{1,2,3}$,
Alex~Andersson$^{4}$,
Zwidofhela~N.~Khangale$^{1,2}$,
\newauthor
Francesco~Cavallaro$^{5}$,
Steven~Bloemen$^{3}$,
Paul~Vreeswijk$^{3}$,
David~A.~H.~Buckley$^{2}$,
Rob~P.~Fender$^{1,4}$,
\newauthor
B.~W.~Stappers$^{6}$,
D.~L.~A.~Pieterse$^{3}$,
and R.~Wijnands$^{7}$,
Laura N. Driessen$^{8}$
\\
$^{1}$Department of Astronomy, University of Cape Town, Private Bag X3, Rondebosch 7701, South Africa\\
$^{2}$South African Astronomical Observatory, P.O. Box 9, Observatory, 7935, South Africa\\
$^{3}$Department of Astrophysics/IMAPP, Radboud University, P.O. Box 9010, 6500 GL, Nijmegen, The Netherlands\\
$^{4}$Astrophysics, Department of Physics, University of Oxford, Denys Wilkinson Building, Keble Road, Oxford OX1 3RH, UK\\
$^{5}$INAF--Osservatorio Astrofisico di Catania, Via S.~Sofia 78, I-95123 Catania, Italy\\
$^{6}$Jodrell Bank Centre for Astrophysics, University of Manchester, Oxford Road, Manchester M13 9PL, UK\\
$^{7}$Anton Pannekoek Institute for Astronomy, University of Amsterdam, P.O. Box 94249, 1090 GE Amsterdam, The Netherlands\\
$^{8}$Sydney Institute for Astronomy, School of Physics, The University of Sydney, New South Wales 2006, Australia
}

\date{Accepted 24 July 2026. Received 23 July 2026; in original form 26 Feb 2026}
 
\pubyear{\the\year{}}

\begin{document}
\label{firstpage}
\pagerange{\pageref{firstpage}--\pageref{LastPage}}
\maketitle

\begin{abstract}
We report the discovery and characterisation of MKT~J032848.4--271904.6\allowbreak, a new radio transient identified in the LADUMA field using SARAO Science Data Processor SDP UHF-band images from approximately one year of MeerKAT observations. Using the Transient Pipeline TraP and advanced filtering techniques, we identified a number of candidate variable radio sources in the field. All but one show variability consistent with refractive interstellar scintillation, leaving MKT~J032848.4--271904.6 \allowbreak as the sole source exhibiting intrinsic variability. In addition, MKT~J032848.4--271904.6 \allowbreak shows intrinsic variability at 0.816 GHz, with 13 radio detections across 41 epochs and a peak flux density of 1.041 $\pm$ 0.043 mJy\allowbreak. We associate this emission with a low-mass M-dwarf star LP~888–63, located 23 pc from the Sun and identified as a companion to the white dwarf binary system LAWD 14 WD~0326--273. LP~888--63 displays active flaring behaviour across the electromagnetic spectrum including multi-band optical data from MeerLICHT. \textit{TESS} photometry reveals a periodic modulation in the blended light curve of 5.780 $\pm$ 0.507 days, consistent with rotational variability of a mid-M dwarf. Archival ESO spectra reveal H$\alpha$ emission, confirming magnetic activity. These findings highlight the capability of MeerKAT’s SDP imaging and facilities like MeerLICHT for real-time detection and characterisation of stellar transients.
\end{abstract}

\begin{keywords}
stars: activity -- stars: flare -- stars: late-type -- radio continuum: stars -- transients.
\end{keywords}

\section{Introduction}
The radio sky hosts a diverse array of transient and variable sources that trace extreme astrophysical processes, from millisecond-duration coherent bursts produced by pulsars and fast radio bursts (FRBs; \citealt{1968Natur.217..709H,2007Sci...318..777L, 2019ARA&A..57..417C}) to long-lived incoherent synchrotron afterglows of gamma-ray bursts and tidal disruption events (\citealt{1997Natur.389..261F,zauderer2011nature, 2019MNRAS.486.2721B, 2024ApJ...973..104D}). Historically, time-domain radio astronomy relied primarily on follow-up observations of transients identified at higher energies (e.g., \citealt{1997Natur.389..261F, 1994IAUC.6006....1S, zauderer2011nature}).

The advent of modern wide-field radio interferometers including MeerKAT \citep{Jonas2009, 2018ApJ...856..180C}, the Australian Square Kilometre Array Pathfinder \citep[ASKAP;][]{2008ExA....22..151J}, and the LOw Frequency Array \citep[LOFAR;][]{2021PASA...38....9H} has transformed time-domain radio astronomy. These instruments combine wide fields of view, high sensitivity, and regular cadences, enabling systematic discovery of transient and variable sources. 
Surveys such as the LOFAR Multifrequency Snapshot Survey \citep[MSSS;][]{2016MNRAS.456.2321S} and the ASKAP Variables and Slow Transients Survey \citep[VAST;][]{2021PASA...38...54M} have revealed phenomena ranging from short-lived coherent bursts to longer-lived incoherent transients, including gamma-ray burst afterglows \citep{Leung_2021} and tidal disruption events \citep{2024ApJ...973..104D}, highlighting the diversity of the transient radio sky.

Radio transients, whether coherent or incoherent, probe extreme astrophysical processes \citep{2015MNRAS.446.3687P}. Coherent sources such as pulsars \citep{1968Natur.217..709H} and fast radio bursts \citep{2007Sci...318..777L, 2019ARA&A..57..417C} are typically detected via high-time-resolution beamformed searches, while incoherent transients such as supernovae and AGN variability, together with stellar flares that may arise from either coherent or incoherent emission processes, are best captured in image-plane surveys. Beamformed methods excel at millisecond-scale events but cover limited sky area, whereas image-plane surveys provide wide sky coverage and multi-epoch baselines, though they face challenges in detecting faint or short-lived signals \citep{2020MNRAS.491..560D, 2022MNRAS.513.3482A}. Together, these complementary approaches have greatly expanded the discovery space for transient events. Despite these efforts, many transients detected in large radio surveys remain unassociated with multiwavelength counterparts or definitive progenitor systems \citep{Hyman_2005, 2016MNRAS.456.2321S, Murphy_2017, 2022MNRAS.512.5037D}. Detection rates suggest that only 1–5$\%$ of the radio sky is variable, with most confirmed sources linked to follow-up observations rather than blind surveys \citep{2016ApJ...818..105M}.

Among the most abundant yet comparatively under-sampled stellar populations are magnetically active M dwarfs, the most common stellar type in the Galaxy \citep{2006AJ....132.2360H,2010AJ....139.2679B,2015ApJ...812....3W}. These low-mass stars exhibit strong magnetic activity, producing energetic flares and radio bursts through both incoherent gyrosynchrotron emission and coherent processes such as plasma emission or electron cyclotron maser emission \citep{1985ARA&A..23..169D, 1994A&A...285..621B}. Such emission probes stellar magnetospheres, particle acceleration, and star–planet magnetic interactions, although no unambiguous radio detection of a star–planet interaction has yet been confirmed observationally. Early observations of late-type flare stars, including UV Ceti \citep{Lovell1963,1969Natur.222.1126L}, demonstrated intense magnetic activity and established an observational connection between chromospheric heating in dMe stars and the occurrence of radio flares, rather than providing direct evidence from hydrogen emission-line diagnostics \citep{1979ApJ...234..579C, 1987ApJ...323..316C}. Subsequent studies detected radio flares from stars such as Proxima Centauri \citep{bastian1990radiostars, VilladsenHallinan2019}, AD Leonis \citep{Slee_Willes_Robinson_2003}, and YZ Canis Minoris \citep{1976ApJS...30...85L}, while more recent wideband time-domain observations have revealed complex coherent burst phenomenology in systems such as AU Microscopii \citep{2020ApJ...905...23Z, 2024A&A...682A.170B}, highlighting the need for wide-field, high-cadence surveys to capture these rare events \citep{Zic2019, 2021NatAs...5.1233C}. Their frequent and energetic flaring also has significant implications for planetary habitability, as intense activity can erode planetary atmospheres \citep{Khodachenko2007,Lammer2007,2017SoPh..292...89P}, and in some cases may be modulated by orbiting planets \citep{2021NatAs...5.1233C}.

Recent wide-field studies \citep{2021MNRAS.502.5438P, 2024MNRAS.529.1258P, 2024PASA...41...84D} have uncovered numerous active M dwarfs. LOFAR monitoring, for instance, identified 19 nearby systems producing bright, highly circularly polarised bursts, possibly linked to magnetic star–planet interactions \citep{2021NatAs...5.1233C}. Additionally, cross-matching the VLA Sky Survey \citep[VLASS;][]{Lacy_2020} and the LOFAR Two-metre Sky Survey \citep[LoTSS;][]{Shimwell2017} revealed that mid-to-late M dwarfs (M4–M7) often deviate from the canonical Güdel–Benz radio/X-ray relation, suggesting a transition from chromospheric flare-driven emission to magnetospheric auroral-type emission \citep{Callingham_2023, Yiu2024}. MeerKAT and MeerLICHT \citep{2025afas.confE.182E} have also contributed to this emerging picture through commensal transient searches. These have led to the discovery of MKT~J170456.2$-$482100, the first MeerKAT transient, detected in the field of the X-ray binary GX~339$-$4 \citep{2020MNRAS.491..560D}, as well as MKT~J174641.0$-$321404, a flaring M dwarf serendipitously identified in a a weekly monitored field \citep{2022MNRAS.513.3482A}. Related studies of radio-active M dwarfs and stellar coherent emitters have since characterised rapidly rotating, magnetically active systems and explored their long-term variability and radio emission mechanisms in a broader context \citep{Chastain_2023, 2025MNRAS.538L..89S, Chastain2025}. The recent detection of a Type~II radio burst from an M dwarf provides the strongest evidence to date for a CME-like event on another star \citep{2025ApJ...988..227C}. The MeerKAT LADUMA survey—"Looking At the Distant Universe with the MeerKAT Array" is primarily designed to trace the evolution of neutral hydrogen over cosmic time \citep{2016mks..confE...4B}. Its long integrations, deep sensitivity, and multi-epoch coverage also make it an exceptional dataset for commensal transient searches, probing variability on timescales from seconds to years.

In this paper, we report the discovery of a radio-flaring M-dwarf in the LADUMA field identified through commensal transient searches
using MeerKAT. By leveraging LADUMA’s deep, multi-epoch observations, this study provides insights into the magnetic activity of M dwarfs while demonstrating MeerKAT’s capability for time-domain astronomy. In Section \ref{Obs} we detail the radio observations and detection methodology.  Sections \ref{Multi} and \ref{OptSpec} presents the multiwavelength follow-up, comprising optical photometry and spectroscopy, radio pulsation searches, UV photometry, and X-ray observations. Sections \ref{Dis} and \ref{con} provide a discussion of the results and concluding remarks.

\section{Observations} \label{Obs}
The data used in this study were obtained from observations conducted as part of the LADUMA survey \citep{2016mks..confE...4B}. These observations were subsequently searched for radio transients within the framework of the ThunderKAT commensal transient search programme \citep{2016mks..confE..13F}. The dataset consists of a single MeerKAT pointing centred at Right Ascension $53.1267^{\circ}$ and Declination $-28.1325^{\circ}$ (J2000).
Observations were carried out between 2 November 2021 and 12 November 2022 using the UHF receiver, covering the frequency range 580–1015~MHz with a fixed central frequency of 816~MHz. Each epoch had an integration time of approximately 5–10~hours, resulting in a total of 48 Science Data Processor (SDP) continuum images (see Table~\ref{tab:observation-log}). The data were processed using the SARAO Science Data Processor pipeline \citep[SDP;][]{Ratcliffe2020}, which operates in three stages: SDPcal, Continuum, and Spectral pipelines.

The SDPcal Pipeline begins within 15 minutes of an observation’s completion. It performs real-time processing to generate radio frequency interference (RFI) flags and derive calibration solutions. In the 32k mode of the MeerKAT correlator, the data are divided into four frequency bands for concurrent processing. This division optimises narrow spectral-line handling and enhances transient-detection precision. The Continuum Pipeline, designed for longer-duration observations, generates continuum images and constructs continuum-subtraction models. Primary beam correction is then applied to refine the flux calibration. Finally, the Spectral Pipeline produces channelised data products, including single-channel FITS images for each of the 32k frequency channels (see \cite{Ratcliffe2020} for more details).

\subsection{The transient pipeline}
Before running the LOFAR Transients Pipeline (TraP; \citealt{swinbank2015lofar, 2019A&C....27..111R, 2020MNRAS.491..560D}), we filtered 48 images for our commensal transient search by visually inspecting the images and assessing their noise levels. Images were rejected if affected by strong RFI, poor calibration, or high RMS noise. Two filtering methods were employed: manual inspection using the Cube Analysis and rendering Tool for Astronomy \citep[CARTA;][]{2021zndo...4905459C} software to visually assess images for irregularities, resulting in two images being flagged and removed. Additionally, a Python-based noise level assessment was conducted, where the RMS background noise of each image was measured and compared to the overall distribution across the full dataset. A 3$\sigma$ cutoff relative to the mean RMS noise (13.96 $\pm$ 0.72 $\mu$Jy) was used to identify images with anomalously high noise levels. Seven additional images exceeded this threshold and were flagged, resulting in 41 images that were run through TraP. This dual approach ensured that the selected images for the transient search were of high quality and minimised false positives.

TraP processes time-series imaging data, performing source detection, association, and light curve construction across multiple epochs. Source detection was performed using an $8\sigma$ threshold above the local noise level, consistent with previous studies \citep{2020MNRAS.491..560D,2022MNRAS.517.2894R, 2022MNRAS.513.3482A, Chastain2025}. This threshold balances sensitivity with the need to minimise false positives. The key settings, summarised in Table~\ref{tab:trap_params}, included the use of \texttt{force\_beam=True} to emphasise unresolved point sources. Additionally, the \texttt{deblend\_nthresh} parameter was adjusted to handle overlapping sources, while \texttt{beamwidths\_limit=1.0} ensured proper source association. 

\begin{table}
\caption{Parameters used when running the TraP pipeline.}
\label{tab:trap_params}
\centering
\begin{tabular}{lc}
\hline
Parameter & Value \\
\hline
\texttt{force\_beam} & True \\
\texttt{extraction\_radius\_pix} & 2303 pixels \\
\texttt{deblend\_nthresh} & 10 \\
\texttt{beamwidths\_limit} & 1.0 \\
\texttt{detection\_threshold} & $8\sigma$ \\
\hline
\end{tabular}
\end{table}

Variability was assessed using the reduced chi-squared statistic ($\eta_{\nu}$) and the modulation index ($V_{\nu}$) as also noted in \cite{swinbank2015lofar} and \cite{2019A&C....27..111R}. The reduced chi-squared statistic is defined as:
\begin{center}
    \begin{equation}
        \eta_{\nu} = \chi^2_{N-1} = \frac{1}{N-1} \sum^N_{i=1}\frac{(F_{\nu,i}-\delta^2_{\nu})^2}{\sigma^2_{\nu,i}}
        \label{eq:eta}
    \end{equation}
\end{center}
where $N$ denotes the number of data points in the light curve, $F_{\nu,i}$ and $\sigma_{\nu,i}$ being the flux and associated uncertainty at some observed frequency $v$ and lastly $\delta_{\nu}$ is the weighted mean flux density. The weights are given by $w_{\nu} = \frac{1}{\sigma^2_{\nu}}$.

The $V_{\nu}$ parameter quantifies the variability of flux measurements by comparing the sample standard deviation to the mean of these measurements. This parameter is given by:
\begin{center}
    \begin{equation}
        V_{\nu} = \frac{S_{\nu}}{\overline{F_{\nu}}} = \frac{1}{\overline{F_{\nu}}} \sqrt{\frac{N}{N-1}(\overline{F^2_{\nu}} - \overline{F_{\nu}}^2)}
    \label{eq:V}
    \end{equation}
\end{center}
where $S_{\nu}$, $\overline{F_{\nu}}$ and $N$ denotes the light curves standard deviation, mean and number of data points in a given light curve respectively. For a given source, we have that a low $V_\nu$ value depict a limited range of flux densities, while a substantial $V_\nu$ value suggests a wide dispersion in flux densities, reflecting heightened variability \citep{2022MNRAS.512.5037D}. Sources with high values of both $\eta_{\nu}$ and $V_{\nu}$ are classified as highly variable or transient. For a more detailed description of the TraP see \cite{swinbank2015lofar}, \cite{2019A&C....27..111R} and \cite{2020MNRAS.491..560D}.

The TraP output included a catalogue of 7\,299 detected sources. To identify genuine transients and variables, we applied filtering based on the variability metrics $\eta_\nu$ and $V_\nu$ \citep{2019A&C....27..111R,2022MNRAS.513.3482A}. We modelled the logarithmic distributions of $\eta_\nu$ and $V_\nu$ as Gaussian and defined statistical outliers as sources lying beyond $2\sigma$ from the mean of each distribution. For the LADUMA field at 816\,MHz, the corresponding thresholds were $\eta_\nu = 275.15$ and $V_\nu = 0.30$.

\begin{figure}
    \centering
    \includegraphics[width=1.0\linewidth]{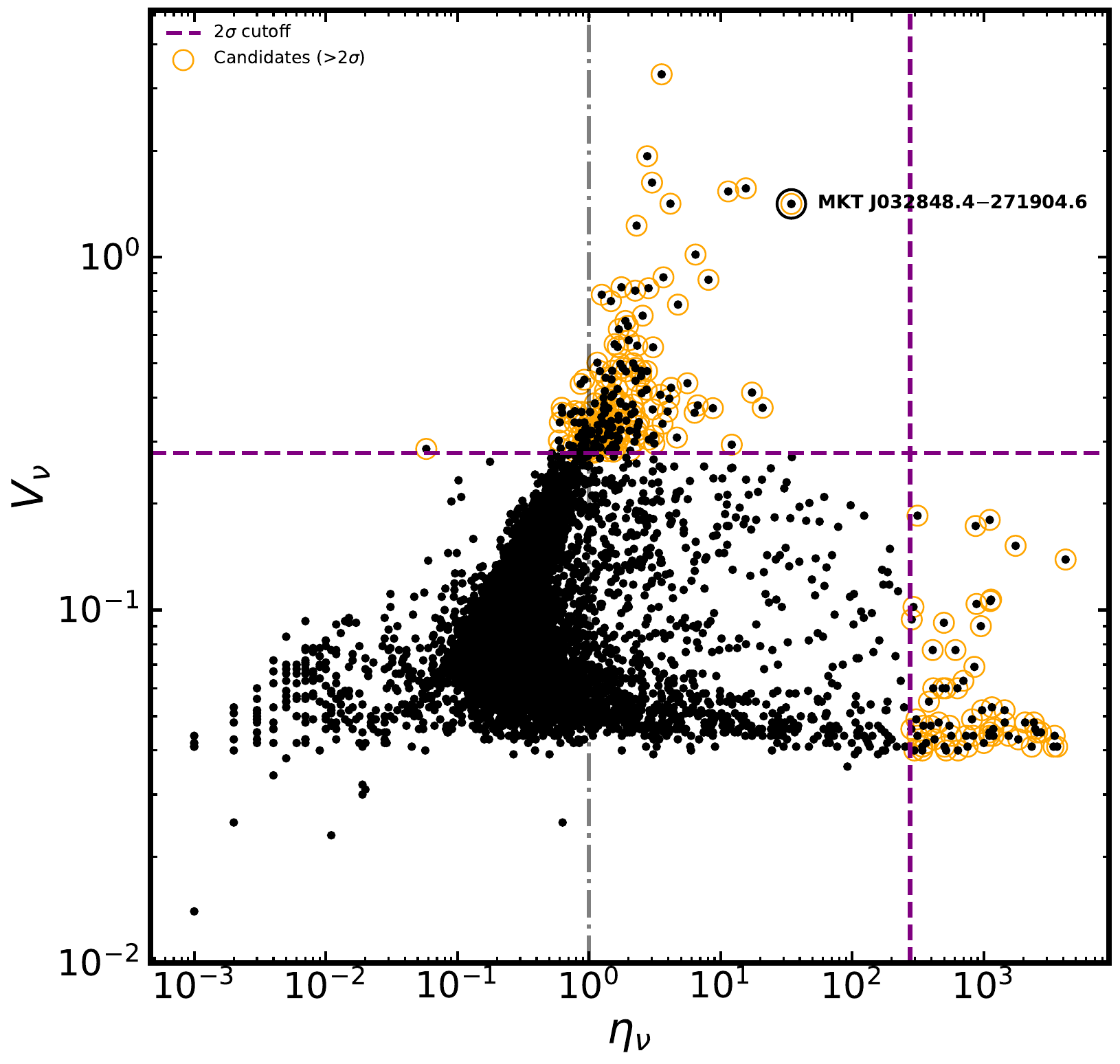}
    \caption{
    Variability parameter space defined by $\eta_\nu$ and $V_\nu$ for all sources detected in the LADUMA field. The dashed lines indicate the $2\sigma$ thresholds derived from the distributions of each statistic. No sources are present in the classical transient quadrant where both thresholds are exceeded simultaneously. Sources individually exceeding either threshold and subsequently investigated are highlighted with orange circles, while MKT~J032848.4--271904.6 is labelled as the only source exhibiting variability consistent with intrinsic behaviour rather than scintillation or imaging artefacts.
    }
    \label{fig:variability}
\end{figure}

No sources were found in the classical transient quadrant defined by simultaneously exceeding the $2\sigma$ thresholds in both $\eta_\nu$ and $V_\nu$ (the upper-right region of parameter). 
We therefore inspected all sources exceeding either threshold individually, noting that variability in only one statistic does not necessarily imply intrinsic transient behaviour (seen in Figure \ref{fig:variability} denoted in circles). The investigated sources predominantly exhibited variability consistent with scintillation or imaging artefacts, with the exception of MKT~J032848.4--271904.6, which is the only source showing behaviour consistent with intrinsic variability and is therefore presented here.

\begin{figure*}
    \centering
    \includegraphics[width=1\linewidth]{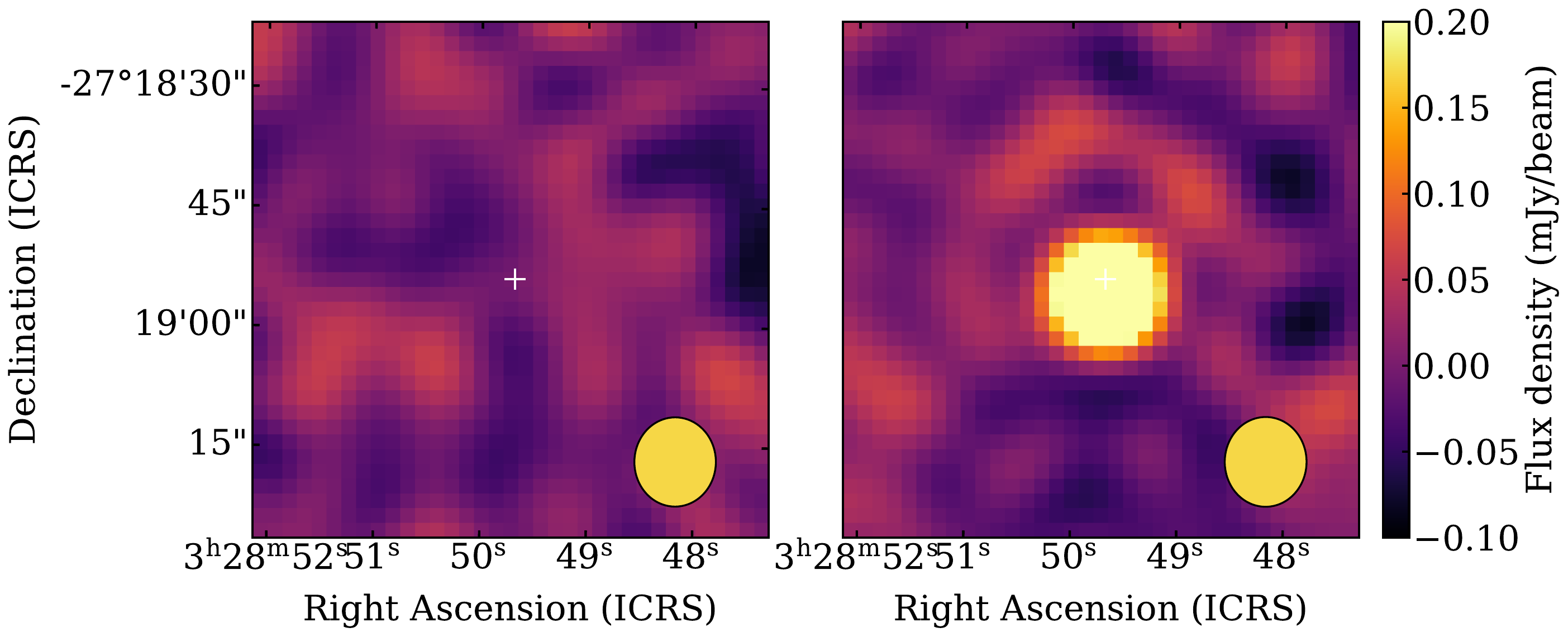}
    \caption{Radio image cut-outs centred on the position of MKT~J032848.4--271904.6 for two representative LADUMA epochs. 
The left panel shows a non-detection, while the right panel displays a clear detection of the source. 
Both images are presented in the ICRS reference frame with flux density shown in mJy\,beam$^{-1}$. 
The synthesised beam is indicated by the ellipse in the lower-right corner of each panel, and the cross marks the target position.
}
    \label{fig:Radio_image}
\end{figure*}

\begin{figure}
    \centering
  \begin{subfigure}{1.0\linewidth}
    \centering
    \includegraphics[width=\linewidth]{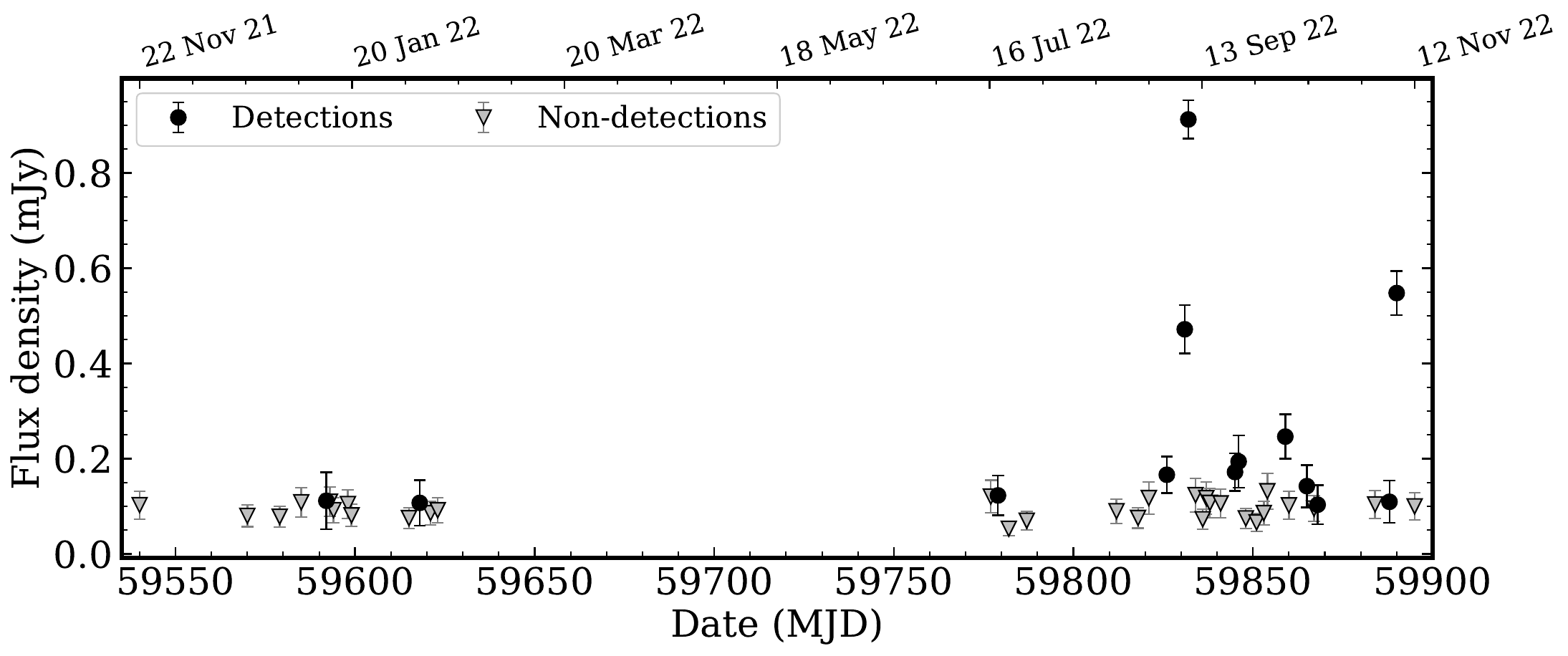}
    \label{fig:FullLight}
  \end{subfigure}
  \hfill
  \begin{subfigure}{1.0\linewidth}
    \centering
    \includegraphics[width=\linewidth]{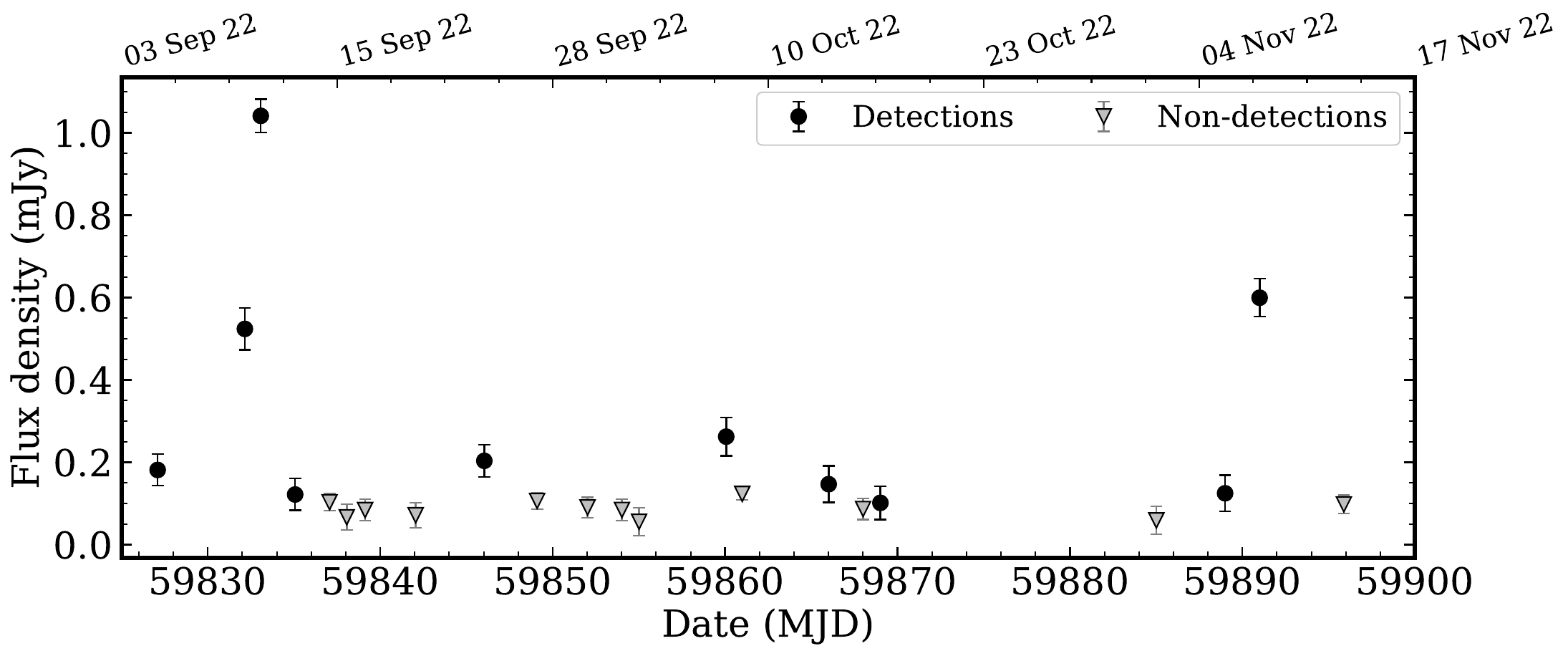}
    \label{fig:ZoomLC}
  \end{subfigure}
   \caption{\textbf{Top:} The radio light curve of MKT J032848.4--271904.6 across all 41 LADUMA images. This shows all 13 detections from TraP monitoring, non-detections were treated as upper limits, defined as flux measurements below the $3\sigma$ threshold, where $\sigma$ represents the local noise level for each epoch. 
   \textbf{Bottom:} shows 10 detections from the second season MeerKAT's observations of the LADUMA field obtained from the TraP running catalogue.}
  \label{fig:LightCurveFigure}
\end{figure}

MKT~J032848.4--271904.6 is a significant outlier in the $\eta_\nu$–$V_\nu$ distribution (Figure~\ref{fig:variability}), with $\eta_\nu = 32.951$ and $V_\nu = 1.372$. Its peak flux density is $1.041 \pm 0.043$\,mJy, and the source is detected in 13 of the 41 observing epochs. Figure~\ref{fig:Radio_image} presents radio images from two representative epochs, showing a non-detection and a clear detection. Non-detections are treated as $3\sigma$ upper limits based on the local image noise in each epoch. The resulting radio light curves from the TraP monitoring and the running catalogue are shown in Figure~\ref{fig:LightCurveFigure}. 
The top panel displays all 13 detections from the TraP monitoring, while the bottom panel shows the 10 detections obtained during the second LADUMA observing season from the TraP running catalogue.

\section{Multiwavelength counterparts} \label{Multi}
The position of MKT~J032848.4$-$271904.6 was refined using PyBDSF \citep{2015ascl.soft02007M}, yielding a radio centroid at RA\,=\,03:28:49.69, Dec\,=\,$-$27:18:55.71. A cross-match with SIMBAD \citep{2000A&AS..143....9W} within 2$^{\prime}$ returns two nearby optical sources: LAWD~14 and LP~888$-$63. The physical and astrometric properties of LP~888$-$63 are summarised in Table~\ref{tab:properties}; values repeated in subsequent sections refer to this table.

\begin{table}
\centering
\caption{Physical and astrometric properties of LP~888$-$63.
Stellar parameters are from \citealt{2019AJ....158..138S}.
Astrometry is from \textit{Gaia}~DR3 \citealt{2021A&A...650C...3G}.
Spectral type and binary membership from \citealt{2005A&A...440.1087N}.
Rotation period and $v_{\rm eq}$ are derived in this work.}
\label{tab:properties}
\begin{tabular}{lr}
\hline
Property & Value \\ \hline
Spectral type                               & M3.5\,V \\
Distance $(\mathrm{pc})$                    & $23.030 \pm 0.020$ \\
RA (ICRS J2016.0)                           & 03:28:48.444 \\
Dec (ICRS J2016.0)                          & $-$27:19:04.603 \\
$\mu_{\alpha*}$ $(\mathrm{mas\,yr}^{-1})$   & $733.085 \pm 0.024$ \\
$\mu_{\delta}$ $(\mathrm{mas\,yr}^{-1})$    & $387.658 \pm 0.035$ \\
Mass $(\mathrm{M}_{\sun})$                  & $0.297 \pm 0.020$ \\
Radius $(\mathrm{R}_{\sun})$                & $0.317 \pm 0.010$ \\
$T_{\mathrm{eff}}$ $(\mathrm{K})$           & $3297 \pm 157$ \\
Rotation period $(\mathrm{d})$              & $5.780 \pm 0.507$ \\
$v_{\mathrm{eq}}$ $(\mathrm{km\,s}^{-1})$  & $2.77 \pm 0.12$ \\
$G$-band magnitude $(\mathrm{mag})$         & $12.430 \pm 0.003$ \\
H$\alpha$ EW (\text{\AA})                   & $-1.74 \pm 0.05$ \\ \hline
\end{tabular}
\end{table}

\begin{figure}
\centering
\includegraphics[width=0.485\textwidth]{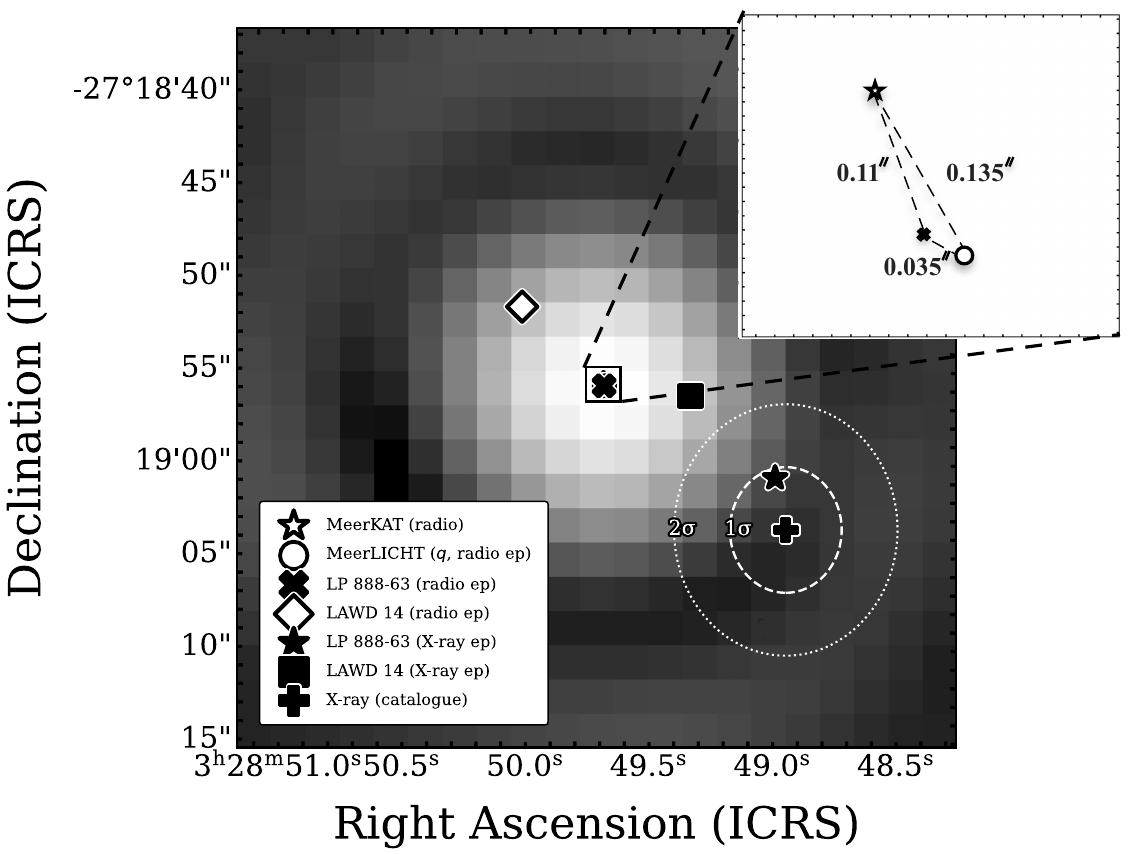}
\caption{Astrometric overlay in the ICRS frame showing the association of MKT~J032848.4$-$271904.6 across radio, optical, and X-ray wavelengths. The MeerKAT continuum image at the radio epoch (2022.676) is shown in greyscale. \textit{Gaia}~DR3 positions of LP~888$-$63 and LAWD~14, together with the MeerLICHT $q$-band mean position, are propagated to the radio epoch using their catalogued proper motions. The inset highlights the radio--optical separations, with LP~888$-$63 lying within 0.11$^{\prime\prime}$ of the radio centroid and LAWD~14 offset by 6.05$^{\prime\prime}$. The \textit{XMM-Newton} Slew Survey position (epoch 2010.05) is shown with its 1$\sigma$ and 2$\sigma$ uncertainty regions; at the X-ray epoch, LP~888$-$63 lies within the 1$\sigma$ region, while LAWD~14 lies outside 2$\sigma$.}
\label{fig:astrometry_overlay}
\end{figure}

\textit{Gaia}~DR3 astrometry is referenced to epoch J2016.0; the MeerKAT image is tied to the ICRS frame at 2022 September~4 (decimal year 2022.676). The catalogued proper motions of LAWD~14 and LP~888$-$63 (Table~\ref{tab:properties}) are used to propagate their \textit{Gaia}~DR3 coordinates to the MeerKAT epoch; \textbf{full astrometric details are given in Table~\ref{tab:astrometry}.} After propagation, LP~888$-$63 lies within 0.11$^{\prime\prime}$ of the MeerKAT radio centroid, while LAWD~14 is offset by 6.05$^{\prime\prime}$ and is ruled out as a counterpart (Figure~\ref{fig:astrometry_overlay}).

To assess the X-ray association, \textit{Gaia}~DR3 positions are propagated to the \textit{XMM-Newton} Slew Survey epoch (2010 January~19; decimal year 2010.05). The catalogue position of XMMSL2~J032848.9$-$271903 has a 1$\sigma$ uncertainty of 3.39$^{\prime\prime}$. At the X-ray epoch, LP~888$-$63 lies 2.86$^{\prime\prime}$ from the X-ray centroid (within 1$\sigma$), while LAWD~14 is offset by 8.87$^{\prime\prime}$ (outside 2$\sigma$). The \textit{Gaia} positional uncertainties at both epochs are $\lesssim$0.03$^{\prime\prime}$ and are negligible compared to the X-ray error budget. Taken together, the consistent positional agreement across radio, optical, and X-ray wavelengths uniquely identifies LP~888$-$63 as the counterpart to MKT~J032848.4$-$271904.6 (Figure~\ref{fig:astrometry_overlay}).

\begin{table*}
\caption{Astrometric summary of MKT~J032848.4$-$271904.6 and associated multi-wavelength sources...}
\label{tab:astrometry}
\centering
\begin{tabular*}{\textwidth}{@{\extracolsep{\fill}} l l l l l l }
\hline
Physical Object & Instrument & Epoch & RA (deg) & Dec (deg) & Separation (arcsec) \\ \hline
LP~888$-$63 (Radio peak)    & MeerKAT             & 2022.676 & 52.20706 & $-27.31548$ & ---                     \\[4pt]
LP~888$-$63 (Optical)       & \textit{Gaia}~DR3   & 2022.676 & 52.20705 & $-27.31550$ & 0.11                    \\[4pt]
LP~888$-$63 ($q$-band mean) & MeerLICHT           & 2022.676 & 52.20704 & $-27.31551$ & 0.14                    \\[4pt]
LAWD~14 (WD~0326$-$273)     & \textit{Gaia}~DR3   & 2022.676 & 52.20694 & $-27.31502$ & 6.05                    \\[6pt]
XMMSL2~J032848.9$-$271903   & \textit{XMM-Newton} & 2010.05  & 52.20397 & $-27.31764$ & 2.86 (LP); 8.87 (LAWD)  \\ \hline
\end{tabular*}
\end{table*}

\subsection{Optical photometry}
To further characterise the nature of LP 888--63, we examined its photometric properties using archival datasets and recent time-domain observations. LP 888--63 has extensive multiwavelength coverage across several surveys, including the AllWISE data release \citep{2014yCat.2328....0C}, the Two-Micron All-Sky Survey \citep[2MASS;][]{2003yCat.2246....0C}, the Luyten Half-Second \citep{1979lccs.book.....L}, and the New Luyten Two-Tenths \citep{1979nltt.book.....L}. It is classified as a low-mass eruptive star \citep{2019AJ....157..234S} and a high proper motion star \citep{2016ApJ...817..112S}. 

LP 888--63 is identified as the M(3.5) companion to the white dwarf binary LAWD~14 (WD~0326--273), forming a triple system \citep{2005A&A...440.1087N}. Physical and astrometric properties are given in Table~\ref{tab:properties}; optical and infrared photometric properties are summarised in Table~\ref{tab:magnitudes}. Near-infrared magnitudes from 2MASS are consistent with a mid M-type spectral classification. Time-domain variability from TESS \citep{2015JATIS...1a4003R} and MeerLICHT \citep{2016SPIE.9906E..64B, Groot_2024} confirms optical flaring behaviour consistent with the magnetic activity commonly observed in M dwarfs. 
\begin{table}
\centering
\resizebox{\columnwidth}{!}{%
\begin{tabular}{cccc}
\hline
\begin{tabular}[c]{@{}c@{}}Band\\ $\lambda_{\mathrm{eff}}\,(\mu\mathrm{m})$\end{tabular} & \begin{tabular}[c]{@{}c@{}}Magnitude\\ $(\mathrm{mag})$\end{tabular} & Spectral type &
  Reference \\ \hline
B(0.45)  & 15.6               &  & \citep{1979nltt.book.....L, 1985ApJS...59..197B} \\
V(0.55) &
  $13.80 \pm 0.031$ &
  M3 &
  \citep{1979nltt.book.....L, 1985ApJS...59..197B} \\
G(0.667) & $12.43 \pm 0.0027$ &  & \textbf{\citep{2020yCat.1350....0G}}             \\
J(1.25)  & $9.59 \pm 0.026$   &  & \citep{2003yCat.2246....0C}            \\
H(1.65)  & $9.003 \pm 0.022$  &  & \citep{2003yCat.2246....0C}            \\
K(2.16)  & $8.771 \pm 0.021$  &  & \citep{2003yCat.2246....0C}            \\ \hline
\end{tabular}%
}
\caption{Magnitudes and spectral types for different bands of the low-mass star LP 888--63}
\label{tab:magnitudes}
\end{table}
The combination of broad spectral coverage and multi-epoch variability studies reinforces LP 888--63 as a magnetically active M dwarf, supporting its association with the detected radio flaring events.
\subsubsection{TESS}
\begin{figure}
  \centering
  \begin{subfigure}{0.95\linewidth}
    \centering
    \includegraphics[width=\linewidth]{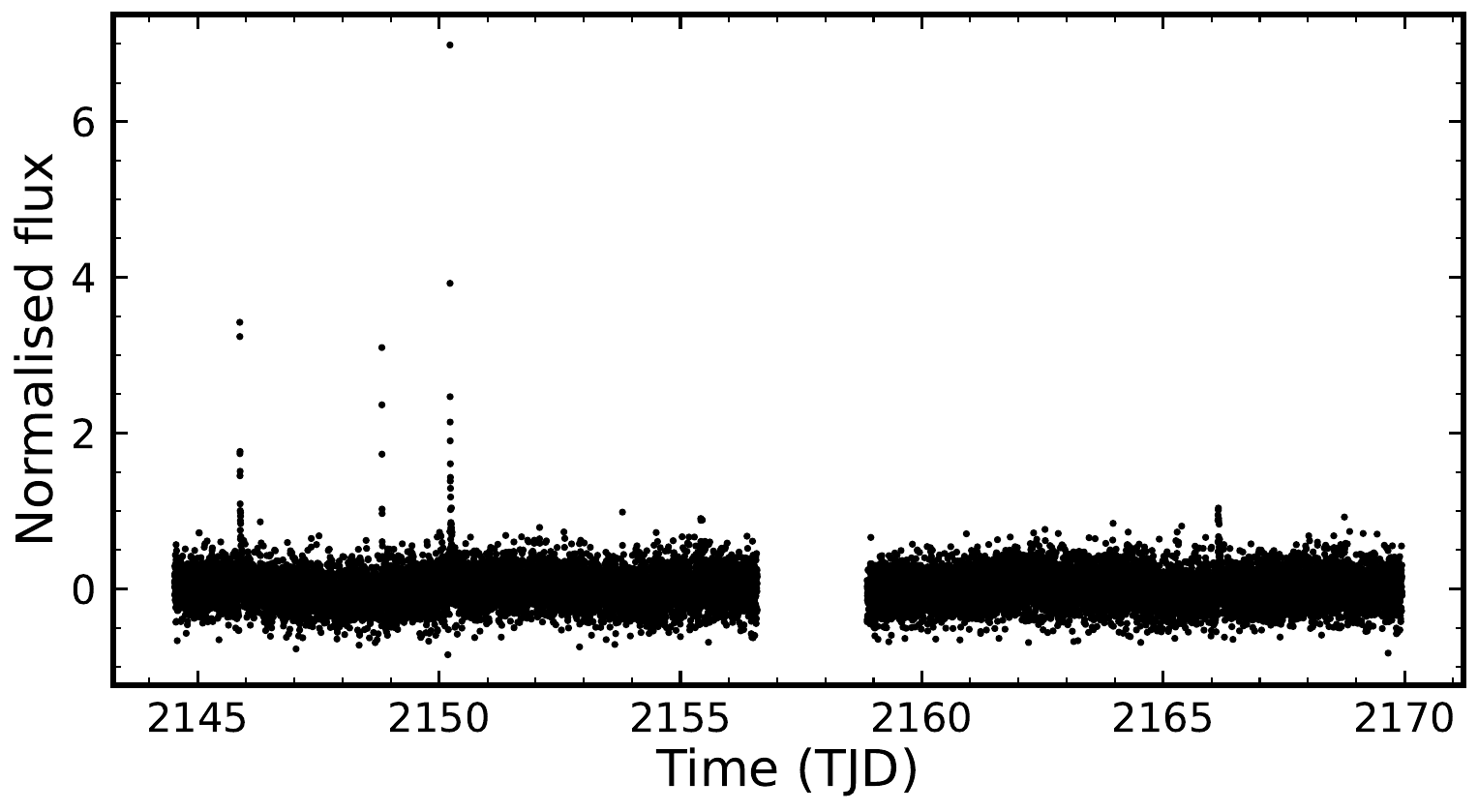}
    \label{fig:TessFullLight}
  \end{subfigure}
  \hfill
  \begin{subfigure}{0.98\linewidth}
    \centering
    \includegraphics[width=\linewidth]{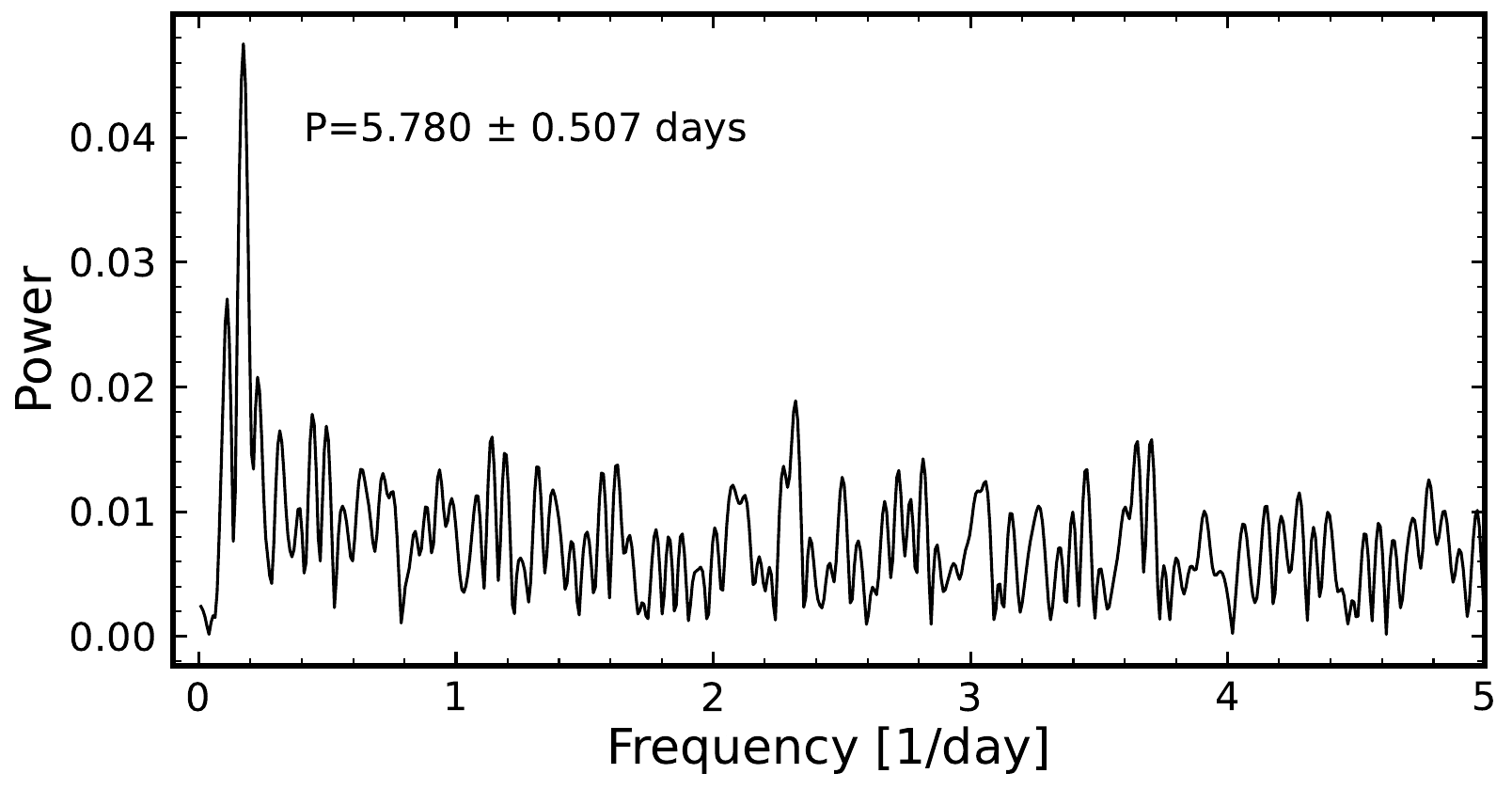}
    \label{fig:PowerPlot}
  \end{subfigure}
  \caption{$\textbf{Top}$ $TESS$ full light curve of LP 888--63 showing clear flares in sector 31. Normalised Flux plotted with time. $\textbf{Bottom}$ Lomb-Scargle periodogram with best frequency of 0.173 days$^{-1}$ and thus period of P $=$ 5.780 $\pm$ 0.507 days. Data shown are from TESS Sector~31 (120-second PDCSAP cadence).}
  \label{fig:overall}
\end{figure}

The source is observed by $TESS$ as TIC 144538915 in both sector 4 and 31 \citep{2019AJ....158..138S}. Clear flares are seen in sector 31 and not in sector 4. The full light curve is shown at top panel of Figure \ref{fig:overall}. We use the 120-second cadence Pre-search Data Conditioning Simple Aperture Photometry (PDCSAP) light curves for both sectors, downloaded from the Mikulski Archive for Space Telescopes (MAST). 
The $TESS$ pixel scale is 21$^{\prime}$, and the angular separation between LP~888–63 and LAWD~14 is significantly smaller than this. The two sources are therefore unresolved in $TESS$, and the extracted light curve represents blended flux from both objects. Consequently, the $TESS$ data cannot distinguish which star dominates a given variability signal, and any periodic modulation derived from these data must be interpreted with this limitation in mind.
We perform a Lomb-Scargle periodogram \citep{1976Ap&SS..39..447L,1982ApJ...263..835S} on the light curve with the aim to determine the periodicity of the star and we get a period of P $=$ 5.780 $\pm$ 0.507 days in sector 31 and P $=$ 6.076 $\pm$ 0.133 days in sector 4. We perform bootstrap analysis to obtain the uncertainties on the period \citep{2014sdmm.book.....I,2018ApJS..236...16V}. Assuming a 63.3 percent confidence interval the values are in agreement as the uncertainties overlap with each other. This is consistent with the value found in sector 4 where a period of P $=$ 6.176 $\pm$ 0.535 days is recorded for this source with two minutes cadence \citep{2023ApJS..268....4F}. Prior to computing the periodogram, identified flare events were masked using an iterative sigma-clipping approach. Removing flares had a negligible effect on the recovered period, confirming that the periodic modulation is dominated by rotational variability rather than stochastic flare activity. Based on visual inspection of the Sector~31 light curve, we identify approximately 4 distinct flare events over the $\sim$27-day sector baseline, corresponding to an estimated flaring rate of $\sim$0.15 flares per day. No clear flares are identified in Sector~4. The source is estimated to have a mass of 0.297$\pm$0.020 $M_{\odot}$, radius of 0.317$\pm$0.010 $R_{\odot}$ and an effective temperature of 3297.0$\pm$157.0 K in The $TESS$ Input Catalog and Candidate Target list \citep{2019AJ....158..138S}. This effective temperature is consistent with values reported by \cite{2005A&A...440.1087N} and \cite{2014AcA....64..359T}. If the detected modulation arises from LP~888–63 and reflects stellar rotation, assuming the $TESS$ radius and our calculated period from the $TESS$ sector 31 light curve, and assuming that the source rotates as a rigid body we use $v_t = 2\pi R / P$ and propagate the uncertainties to calculate the rotational velocity of the source to be 2.77$\pm$0.12 km s$^{-1}$. This corresponds to relatively slow stellar rotation, comparable to the solar equatorial velocity, and is consistent with a mid-M dwarf exhibiting a multi-day rotation period, as observed in nearby mid-M dwarf populations \citep{Newton2016}. However, because the $TESS$ photometry is blended, this inferred rotational velocity should be regarded as tentative. We emphasise that this rotational velocity should not be confused with tangential space velocities used in stellar kinematics, which are typically tens of km s$^{-1}$ for nearby dwarf stars.

\subsubsection{MeerLICHT} \label{MeelL}
The optical counterpart LP~888$-$63 was observed with the MeerLICHT telescope as part of routine monitoring during MeerKAT radio observations, providing contemporaneous multi-band optical coverage in the $u g q r i z$ filters.

\begin{figure}
    \centering
    \includegraphics[width=1\linewidth]{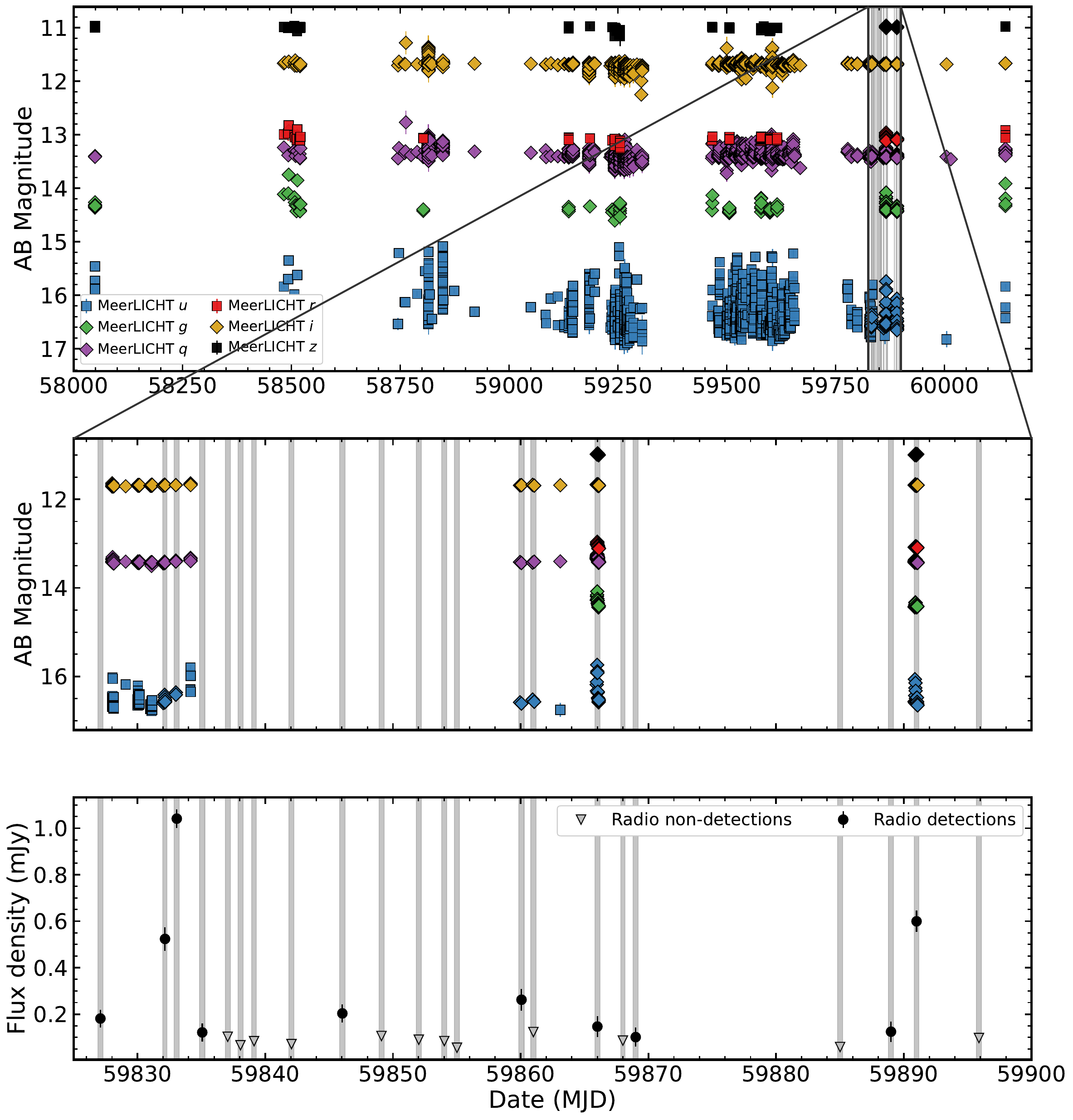}
    \caption{Multi-wavelength light curve of MKT~J032848.4--271904.6. \textbf{Top:} Full optical light curve from MeerLICHT in the $u$, $g$, $q$, $r$, $i$, and $z$ bands plotted in AB magnitude as a function of Modified Julian Date. \textbf{Middle:} Zoomed view of the optical light curve over the interval containing the radio activity, with shaded vertical bands indicating the epochs of radio observations. \textbf{Bottom:} Radio flux-density measurements over the same time range, showing detections and non-detections in mJy. Together, the panels show the temporal relationship between the optical variability and the radio emission. 
}
    \label{fig:meerlicht_uband}
\end{figure}

\textbf{Optical variability and flare phenomenology:}
The source exhibits persistent optical variability across all MeerLICHT bands, with frequent impulsive brightenings characteristic of stellar flares. The most pronounced flaring activity is detected in the $u$ band (Figure~\ref{fig:meerlicht_uband}), with decreasing amplitudes toward longer wavelengths. This wavelength-dependent behaviour is a well-established signature of flares on active M dwarfs, arising from rapid energy deposition in the lower stellar atmosphere that produces a hot continuum component peaking in the blue and near-ultraviolet \citep[e.g.][]{2013ApJS..207...15K,2014ApJ...797..122D}. Because the MeerLICHT observations span multiple observing seasons separated by long gaps, flare rates are normalised by the cumulative on-source monitoring time rather than the total calendar duration of the campaign. In the MeerLICHT $u$ band, we identify 43 distinct flare events over 7.14 days of monitoring, corresponding to a flare rate of $\sim$6.02 flares day$^{-1}$. For comparison, visual inspection of the $TESS$ light curves identifies four flare events in Sector~31 and none in Sector~4, yielding a combined flare rate of $\sim$0.08 flares day$^{-1}$ over the 52.42-day observing baseline. The $u$-band flare rate is therefore approximately 79 times higher than that measured with $TESS$. This is expected, as the MeerLICHT $u$ band is considerably more sensitive to the hot blue continuum produced during stellar flares than the broad $TESS$ bandpass, while differences in cadence and observing strategy also contribute to the observed rate difference.

\begin{figure}
    \centering
    \includegraphics[width=1\linewidth]{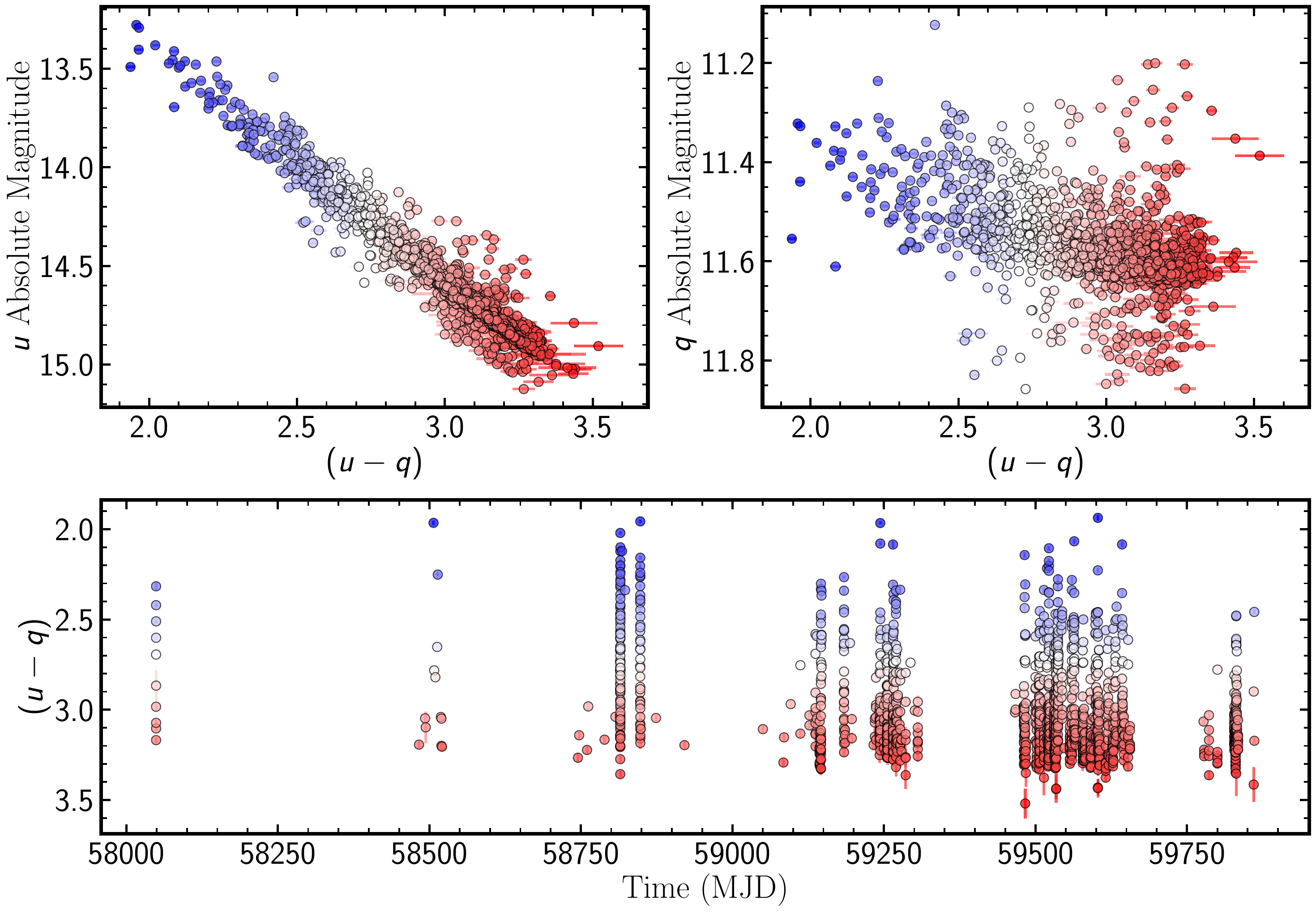}
    \caption{Colour--magnitude and colour--time behaviour of LP~888$-$63 across all MeerLICHT observations. \textbf{Top left:} $u$-band absolute magnitude as a function of colour $(u-q)$. \textbf{Top right:} $q$-band absolute magnitude as a function of $(u-q)$. \textbf{Bottom panel:} temporal evolution of the colour $(u-q)$ as a function of Modified Julian Date. Note that the colour axis $(u-q)$ in the bottom panel corresponds directly to the horizontal axis in the upper panels.}
    \label{fig:meerlicht_cmd}
\end{figure}

The colour--magnitude behaviour of LP~888$-$63 is shown in Figure~\ref{fig:meerlicht_cmd}. During flaring episodes, the source departs significantly from its quiescent locus, exhibiting a pronounced blueward excursion. Such colour evolution is consistent with the emergence of a hot, white-light flare continuum, often characterised by an apparent colour temperature of $\sim$9000–10\,000~K in M-dwarf flares \citep{2016ApJ...820...95K, 2017ApJ...851...91N}. These characteristics support a flare origin driven by magnetic reconnection and particle acceleration rather than rotational modulation or long-term stellar variability.

\begin{figure}
    \centering
    \includegraphics[width=1\linewidth]{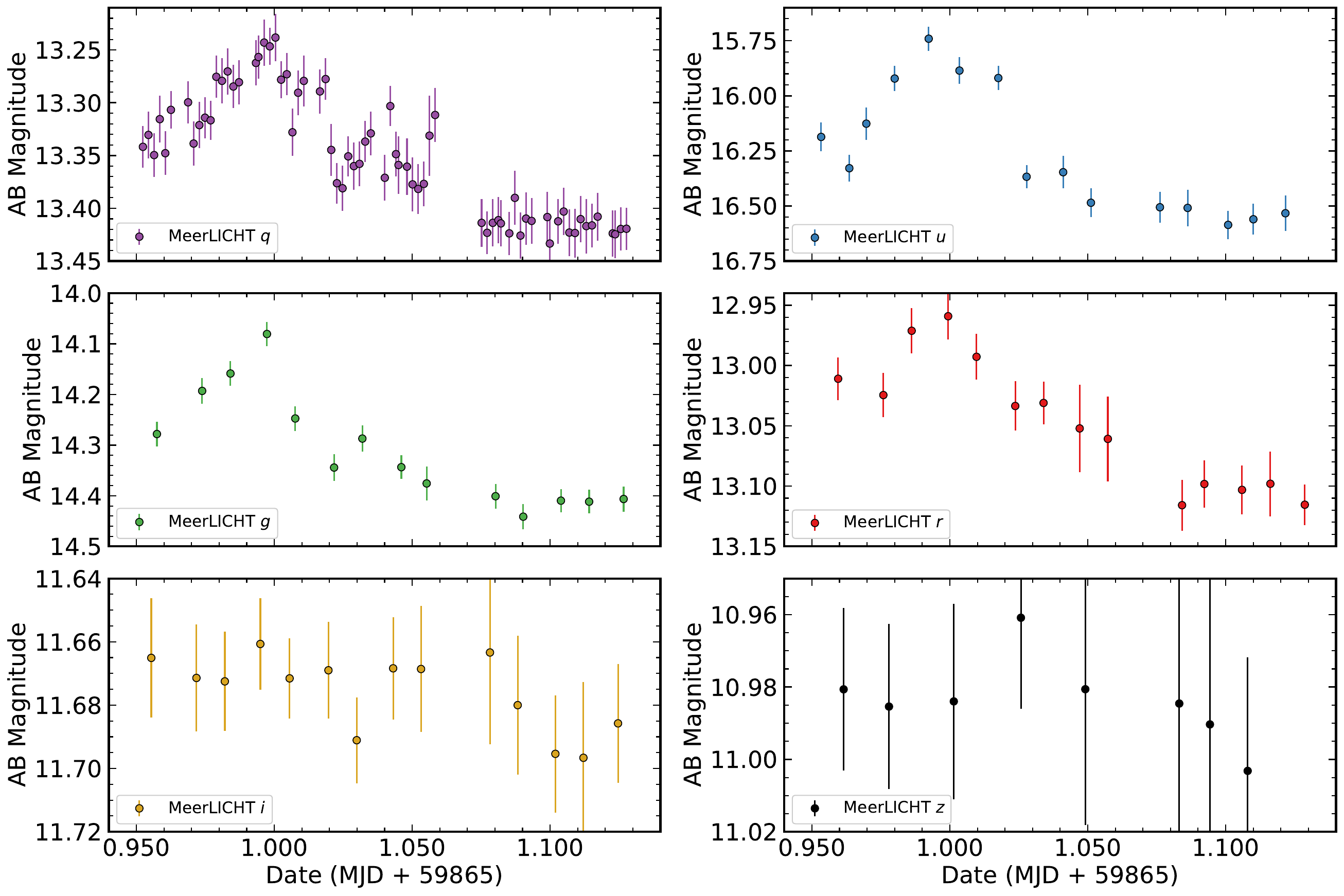}
    \caption{MeerLICHT multi-band light curves of LP~888--63 zoomed in on a single flare event. A clear rise and subsequent decay is observed in the $u$, $g$, $r$, and $q$ filters, while the $i$ and $z$ bands remain comparatively flat within the photometric scatter.}
    \label{fig:meerlicht_zoom}
\end{figure}

A representative optical flare is shown in Figure~\ref{fig:meerlicht_zoom}, where a clear rise and subsequent decay are observed in the $u$, $g$, $r$, and $q$ bands, while the $i$ and $z$ bands remain comparatively flat within the photometric scatter. The measured indices evolution during this event ($u-g = 2.00 \pm 0.35$~mag, $u-q = 2.98 \pm 0.34$~mag, $r-i = 1.37 \pm 0.13$~mag, $i-z = 0.67 \pm 0.23$~mag) are consistent with a rapid blueward excursion relative to the quiescent state, as expected for impulsive white-light flares on M dwarfs \citep{2013ApJS..207...15K,2014ApJ...797..122D}.

\textbf{Optical–radio overlap and luminosity comparison:}
To enable a physically meaningful comparison between optical and radio activity, we isolate the optical flare component from the underlying quiescent stellar emission and adopt a window-based definition of simultaneity. In this framework, optical and radio activity are considered simultaneous if optical flare emission occurs within the same MeerKAT observing window as a radio detection or non-detection, reflecting the time-averaged nature of the radio measurements.

For each MeerLICHT filter, the light curves were first converted from magnitudes to linear flux density units, and all subsequent analysis was performed in flux space. A robust quiescent baseline was then estimated using a median-based approach with iterative outlier rejection, minimising contamination from flares while accounting for slow quiescent variability and photometric scatter. The optical flare signal was defined as the excess flux above this quiescent baseline, and only statistically significant positive excursions in flux were classified as flare activity. Working in flux space ensures that flare amplitudes and energetics are treated linearly and avoids the non-linear distortions inherent to magnitude units. Baseline-subtraction techniques of this form are widely used in optical flare studies and have been shown to robustly recover flare energetics and colour evolution across a range of cadences and instruments \citep[e.g.][]{2016ApJ...829...23D, 2020AJ....160..219F, 2020AJ....159...60G}.

MeerKAT observations of LP~888$-$63 were conducted in discrete start–stop observing blocks, each treated here as a single radio observing window. The measured radio flux density represents an average over the full duration of each window and is plotted at the temporal midpoint for visualisation; consequently, sub-window timing information and instantaneous radio flare peak times cannot be constrained with these data.

Figure~\ref{fig:meerlicht_uband} (middle and bottom panel) presents the full MeerLICHT optical light curves of LP~888$-$63 with MeerKAT observing windows indicated, together with the corresponding MeerKAT radio detections and $3\sigma$ upper limits. Across the monitoring campaign, six MeerKAT observing windows contain simultaneous MeerLICHT observations, permitting a direct assessment of optical–radio co-occurrence on epoch timescales. The optical behaviour within these windows is heterogeneous, ranging from pronounced multi-band optical flaring above quiescence to intervals consistent with quiescent emission despite radio detection.

\begin{figure}
\centering
\includegraphics[width=1\linewidth]{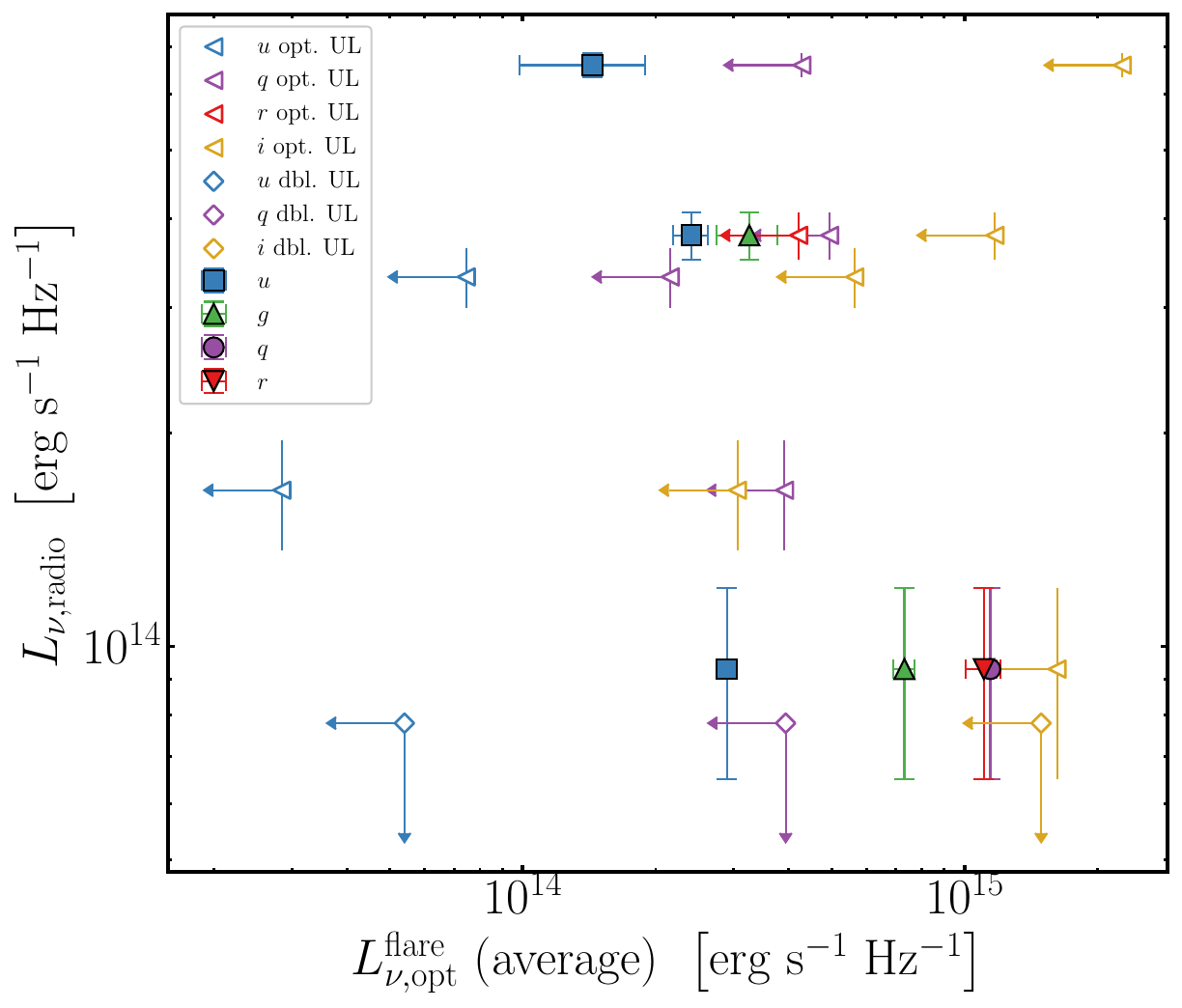}
\caption{Window-averaged radio luminosity versus window-averaged optical flare luminosity for LP~888$-$63. Points correspond to MeerKAT observing windows with contemporaneous MeerLICHT coverage, with optical flare luminosities measured relative to the quiescent baseline and classified as significant at $>3\sigma$. Filled markers denote windows with both a radio detection and a significant optical flare excess. Left-pointing open triangles denote windows with a radio detection but no significant optical flare excess, plotted at their $3\sigma$ optical flare-luminosity upper limit. Open diamonds denote windows with neither a radio detection nor a significant optical flare excess, plotted with both a $3\sigma$ radio upper limit (downward arrow) and a $3\sigma$ optical upper limit (leftward arrow). Different colours indicate optical filters.}
\label{fig:radio_optical_lum_window}
\end{figure}

To quantify the relationship between optical and radio activity in a manner consistent with the time-averaged nature of the radio measurements, we compute window-averaged radio luminosities and corresponding optical flare luminosities for each overlapping window. Optical flare luminosities are derived exclusively from the flare-only signal relative to the quiescent level, ensuring that only impulsive emission contributes to the comparison. The resulting window-averaged radio versus optical flare luminosities are shown in Figure~\ref{fig:radio_optical_lum_window} 

As a supplementary test, we also examine whether the most optically energetic flares above quiescence preferentially occur during radio-bright epochs by comparing the peak optical flare luminosity within each window to the window-averaged radio luminosity. This comparison is shown in Figure~\ref{fig:app_peak_lum} (Appendix \ref{Append}). Since this diagnostic compares a peak optical quantity to a time-averaged radio measurement, it is included for completeness and is not used to define the primary conclusions of this study. The full set of overlapping observing windows is shown in Appendix \ref{Append}, where representative examples of the window-level optical and radio behaviour relative to quiescence are presented.

\subsection{X-ray observations}

The source XMMSL2~J032848.9$-$271903 was detected in the \textit{XMM-Newton} Slew Survey (XMMSL2) with a reported X-ray flux of $1.4 \times 10^{-12}$~erg~s$^{-1}$~cm$^{-2}$, corresponding to an X-ray luminosity of $8.86 \times 10^{28}$~erg~s$^{-1}$ at the distance of LP~888$-$63. To investigate the relationship between the radio and X-ray emission from this source, we place the radio and X-ray luminosities of LP~888–63 in the broader context of the empirical radio–X-ray luminosity plane for magnetically active stars \citep{1994A&A...285..621B}, which links non-thermal radio emission from accelerated electrons to thermal coronal X-ray emission. More recent work has shown substantial scatter and deviations from this relation during flaring states \citep[e.g.][]{2025MNRAS.540.2685E}, reinforcing the need for caution when comparing non-simultaneous measurements. 

As the X-ray and radio observations are not contemporaneous, the comparison is necessarily indicative rather than definitive. In particular, the quiescent radio emission from LP~888–63 is not directly detected. Instead, we estimate an upper limit on the quiescent radio flux density. Specifically, we adopt three times the minimum local RMS noise measured across the 41 individual MeerKAT images as a conservative upper limit on the quiescent emission, rather than stacking non-detection epochs. This gives a quiescent radio flux density limit of 12~$\mu$Jy (Table~\ref{tab:observation-log}), corresponding to a radio luminosity upper limit of $2.27 \times 10^{13}$~erg~s$^{-1}$~Hz$^{-1}$. At peak brightness, the radio luminosity reaches $6.59 \times 10^{14}$~erg~s$^{-1}$~Hz$^{-1}$, exceeding the quiescent level by more than an order of magnitude.

\begin{figure}
    \centering
    \includegraphics[width=0.9\linewidth]{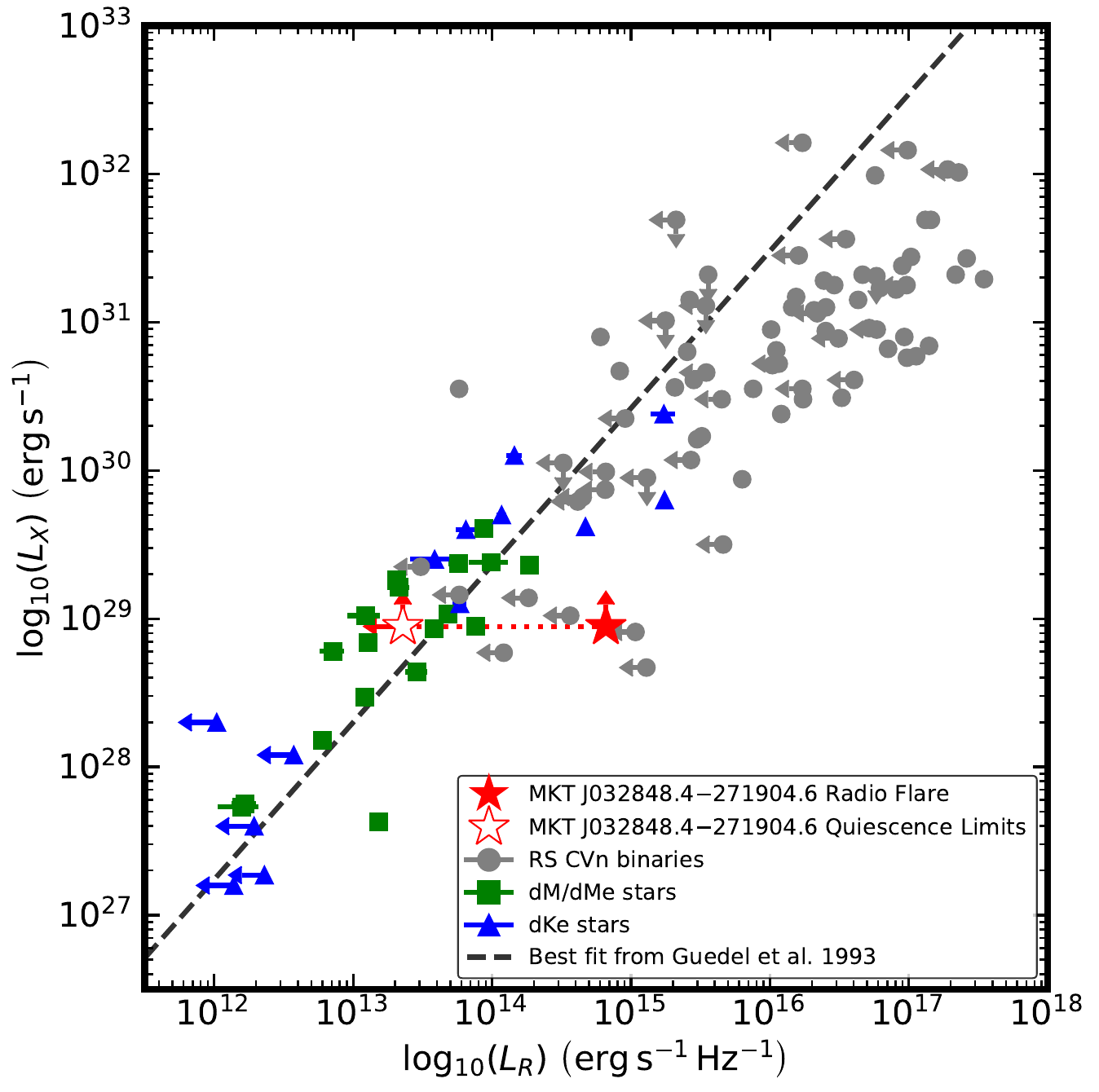} 
    \caption{Radio versus X-ray luminosities of active stars from \citet{1994A&A...285..621B}. The x-axis represents radio luminosity, while the y-axis represents X-ray luminosity. The plot includes various active stellar types, such as RS CVn binaries, M dwarfs (DMe), and K-dwarfs (dKe). The limits on quiescent radio emission for LP~888--63 are shown, along with its peak flare luminosity, which reaches \(6.59 \times 10^{14}\)~erg~s\(^{-1}\)~Hz\(^{-1}\). Adapted from \protect\url{https://github.com/AstroLaura/GuedelPlot}.}
    \label{fig:Radio X-ray}
\end{figure}

To visualise the relative location of LP~888--63 in the radio--X-ray luminosity plane, Figure \ref{fig:Radio X-ray} shows the empirical distribution of magnetically active stars compiled by \citet{1994A&A...285..621B}. The peak radio flare luminosity of LP~888$-$63 lies well above the locus of quiescent active M dwarfs in the G\"{u}del--Benz plane, consistent with a transient departure driven by impulsive particle acceleration. Because the available X-ray measurement represents quiescent emission and does not coincide temporally with the radio flaring epochs, this comparison is indicative rather than a direct test of the G\"{u}del--Benz relation in the flaring state. The quiescent radio luminosity upper limit places LP~888$-$63 within the typical locus of active M dwarfs. Simultaneous radio and X-ray observations during a flaring episode would be required to assess whether LP~888$-$63 conforms to or deviates from the empirical relation during active states \citep[e.g.][]{2025MNRAS.540.2685E}.

\subsection{Optical Spectroscopy} \label{OptSpec}

\begin{figure}
    \centering
    \includegraphics[width=1.0\linewidth]{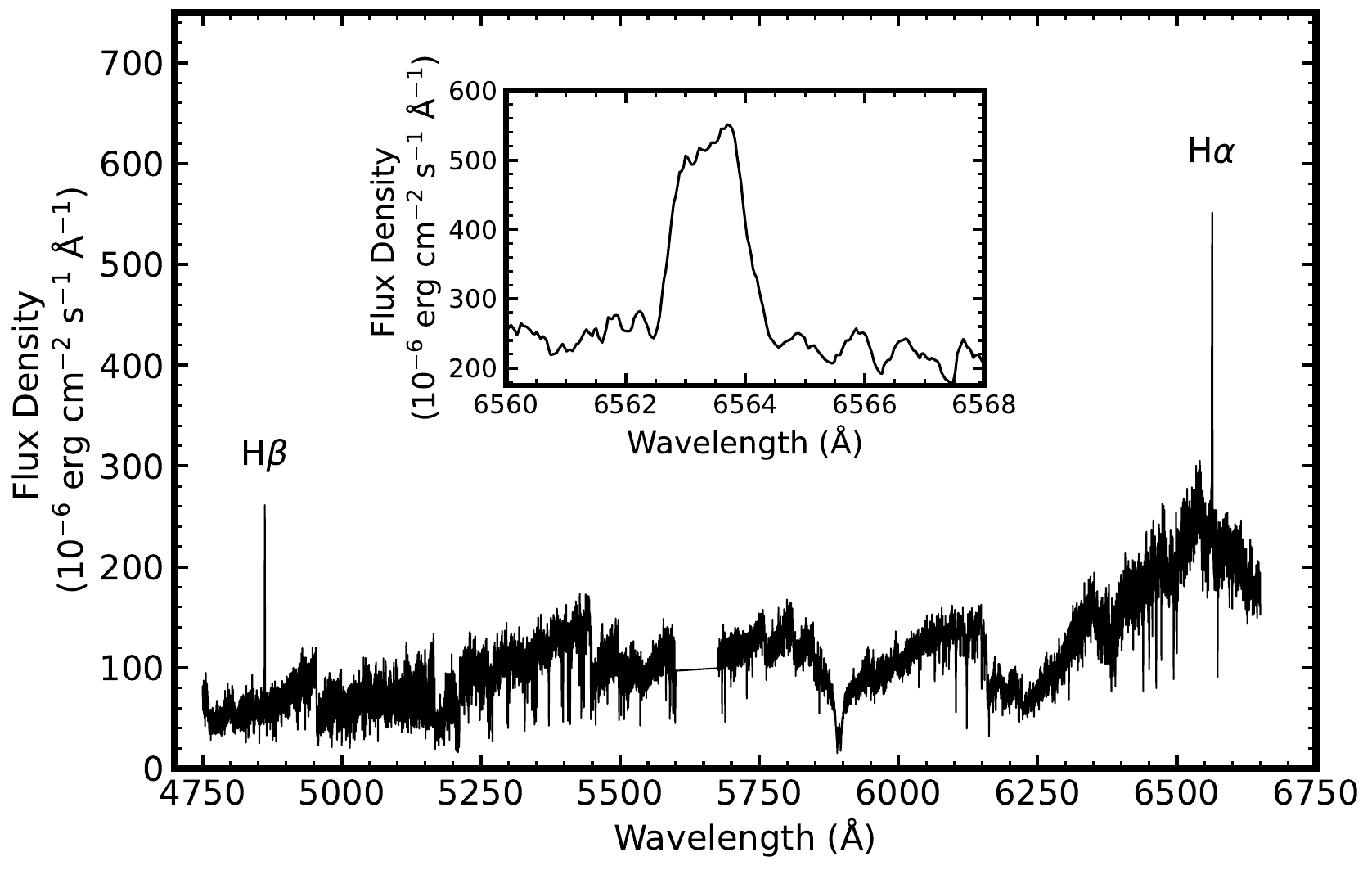}
    \caption{Spectrum of LP 888--63 (L 587--77 B) from the ESO archive. The spectra shows two clear emission lines, H$\alpha$ and H$\beta$. We also show a zoomed view of the H$\alpha$ emission line}
    \label{fig:Spectra}
\end{figure}

The associated optical counterpart LP~888--63 was serendipitously observed during the European Southern Observatory (ESO ESO-VLT-U2) ESO SPY project \cite{2005A&A...440.1087N}. The observations of the target were carried out on 2000-07-12 from 09:57:31 to 10:02:31 (UT), with a total effective exposure time of 300~s using the VLT-UVES high-resolution spectrograph\footnote{\url{http://archive.eso.org/dataset/ADP.2020-09-11T05:49:55.279} (ESO Archive)}. With the spectrum available, we can investigate the spectral type of the star and magnetic activity of the star. The spectrum seen in Figure \ref{fig:Spectra} shows two clear emission lines, H$\beta$ at $\sim$ ${4861}\textup{~\AA}$ and H$\alpha$ at $\sim$ ${6560}\textup{~\AA}$, confirming chromospheric activity. We then calculate the equivalent widths (EW) of the H$\alpha$ emission line integrated between the H$\alpha$ continuum region ${6558.8}\textup{~\AA}$ and ${6566.8}\textup{~\AA}$ detailed by \cite{Newton2017}. We find the calculated EW for the H$\alpha$ to be ${-1.74 \pm 0.05}\textup{~\AA}$, indicating that LP 888--63 is a magnetically active star, as seen in \cite{Newton2017} where stars with EW $<$ ${-1}\textup{~\AA}$ are considered to be magnetically active.

\section{Discussion} \label{Dis}
MKT~J032848.4--271904.6 is spatially coincident with the nearby M3.5\,V star LP~888--63. The source exhibits 13 detections across 41 epochs, with a peak flux density of $1.041 \pm 0.043$~mJy and clear multi-epoch variability (Figures~\ref{fig:Radio_image}~and~\ref{fig:LightCurveFigure}).

Using the observed peak radio flux $f_{v}$, the distance d to the source in parsecs, the observing frequency $v$ in GHz, the characteristic length scale $\text{R}_\text{em}$, and $\text{R}_\text{jup}$ (the radius of Jupiter) we calculate the brightness temperature $\text{T}_{b}$, which provides a diagnostic of the underlying emission mechanism:

\begin{equation}
T_{b}
= 2\times10^{9}
\left(\frac{f_{\nu}}{\mathrm{mJy}}\right)
\left(\frac{\nu}{\mathrm{GHz}}\right)^{-2}
\left(\frac{\text{d}}{\mathrm{pc}}\right)^{2}
\left(\frac{R_\text{em}}{R_\text{jup}}\right)^{-2}
\,\mathrm{K}.
\label{eq:brightness_temperature}
\end{equation}

Following the approach of \citet{2005ApJ...626..486B} and \citet{2022MNRAS.513.3482A}, we initially adopt $\text{R}_\text{em}=2\text{R}_{\star}$ as a conservative upper bound on the characteristic source size. For LP~888--63 we take $\text{R}_{\star} \simeq 0.317\pm0.010\,R_{\odot}$ \citep{2021arXiv210804778P}, yielding a brightness temperature $\text{T}_{b} \sim 4.17\times10^{10}$~K. However, as pointed out by \citet{2022MNRAS.513.3482A}, several studies motivate the use of smaller source sizes:
\citet{VilladsenHallinan2019} adopt $1R_{\star}$ (disk-averaged), while \citet{Yiu2024} explore the range $0.1$--$1R_{\star}$ for coherent bursts. For incoherent flares, source sizes of approximately $1R_{\star}$ have been inferred for very large events \citep{Osten2005} and $\leq0.5R_{\star}$ for moderate flares \citep{Tristan2025}. Adopting $R_\text{em} = 1R_{\star}$ raises $T_b$ to $\sim1.67\times10^{11}$~K, while $R_\text{em} = 0.5R_{\star}$ yields $T_b \sim 6.68\times10^{11}$~K, both approaching the canonical incoherent limit of $\sim10^{12}$~K and placing the emission in an ambiguous regime where coherent processes cannot be excluded on the basis of brightness temperature alone. This ambiguity further motivates the need for circular polarisation (Stokes~$V$) measurements and dynamic spectra to definitively distinguish between gyrosynchrotron and coherent electron cyclotron maser instability (ECMI) emission, as discussed in Section~\ref{Dis}.

From the H$\alpha$ profile we can measure an emission-line FWHM; however, for active M dwarfs this line is chromospheric and its width is dominated by non-rotational broadening (magnetic/Zeeman, Stark, turbulence), so it is not a reliable $v\sin i$ diagnostic \citep[e.g.][]{ReinersBasri2006,Reiners2007}. Instead, we compare the photometric rotation period to the expected equatorial speed, $v_{\rm eq}=2\pi R_\star/P_{\rm rot}$. If the $\sim$6-day periodic modulation detected in the blended \textit{TESS} light curve originates from LP~888--63 and reflects stellar rotation, adopting $\text{R}_\star \simeq 0.317\,R_\odot$ and $\text{P}_{\rm rot}=5.78$~d yields $v_{\rm eq}\approx2.8$~km\,s$^{-1}$, implying $v\sin i\leq2.8$~km\,s$^{-1}$, consistent with a normally rotating, magnetically active mid–M dwarf \citep{Newton2017}. 
Clear optical flares are detected in the $u$, $q$, and $i$ bands, with the source particularly active in $u$, consistent with its prior UV~Ceti–type classification \citep{2014AcA....64..359T}. 

\subsection{Optical--radio flaring behaviour of LP~888--63}

The simultaneous MeerLICHT and MeerKAT monitoring of LP~888--63 enables an investigation of the relationship between optical and radio activity on epoch timescales that are consistent with the cadence and time-averaged nature of the radio observations. By isolating the impulsive optical flare component relative to quiescence and adopting a window-based definition of simultaneity, we assess optical--radio co-occurrence without making assumptions about sub-window timing or instantaneous flare peak alignment.

The MeerLICHT light curves reveal frequent optical flaring, most prominently in the $u$ band, with systematically decreasing amplitudes toward longer wavelengths (Figure~\ref{fig:meerlicht_uband} top panel). This behaviour is characteristic of white-light flares on active M dwarfs and reflects the presence of a hot continuum component produced during impulsive magnetic reconnection and chromospheric heating \citep{2013ApJS..207...15K, 2014ApJ...797..122D, 2016ApJ...820...95K}. The associated colour--magnitude evolution (Figure~\ref{fig:meerlicht_cmd}), marked by pronounced blueward excursions relative to the quiescent locus, is consistent with transient increases in the effective temperature of the emitting region to $\sim$9000--10\,000~K, as inferred from spectroscopic and high-cadence photometric flare studies \citep{2016ApJ...820...95K, 2017ApJ...851...91N, 2020ascl.soft09003H}. These properties place LP~888$-$63 firmly within the population of magnetically active flare stars. The optical flare luminosities observed in LP~888$-$63 are typical of active mid-to-late M dwarfs \citep{2016ApJ...829...23D, 2022ApJ...935..143P} and do not indicate unusually extreme optical activity.

When examined jointly with the MeerKAT observations, a heterogeneous pattern emerges. Figure~\ref{fig:meerlicht_uband} and the window-level examples presented in Appendix~A show that radio detections occur during both optically flaring and optically quiescent intervals. The single radio non-detection window with simultaneous optical coverage (Window 36) was also optically quiescent in every observed filter (Figure \ref{fig:app_window_36}); no case of a significant optical flare coinciding with a radio non-detection was recorded in the present dataset. This diversity indicates that optical and radio activity in LP~888$-$63 are not uniquely coupled on epoch timescales, though the sample of radio non-detection windows with optical coverage remains small. Similar behaviour has been reported in recent radio surveys and coordinated campaigns targeting active M dwarfs, where radio emission may occur independently of detectable optical flaring, and vice versa \citep[e.g.][]{VilladsenHallinan2019, Yiu2024}. The window-averaged radio versus optical flare luminosity comparison (Figure~\ref{fig:radio_optical_lum_window}) provides quantitative summary of this behaviour. The diagram does not show a monotonic or one-to-one relationship between radio luminosity and optical flare luminosity.

Physically, this behaviour is expected given the distinct emission mechanisms involved. Radio emission from active M dwarfs may arise from incoherent gyrosynchrotron radiation produced by mildly relativistic electrons spiralling in coronal magnetic fields or from coherent processes such as plasma emission or ECMI. For incoherent emission, coordinated radio and optical campaigns have observed simultaneous flares with inferred source sizes of order $1R_{\star}$ for large events and $\leq0.5R_{\star}$ for moderate flares \citep{Osten2005, Osten2026, Tristan2025}. For coherent bursts, strong circular polarisation and high brightness temperatures are characteristic signatures, and optical counterparts are not always detected \citep{VilladsenHallinan2019, Yiu2024, Lu2025}. Optical flares primarily probe rapid chromospheric and photospheric heating associated with impulsive energy deposition, whereas the radio luminosities measured with MeerKAT represent time-averaged emission that may arise from incoherent gyrosynchrotron radiation or coherent electron cyclotron maser (ECM) emission in the corona or magnetosphere \citep{1985ARA&A..23..169D, 2002ARA&A..40..217G}. In particular, coherent ECM emission is highly beamed and strongly dependent on magnetic field geometry, and need not be accompanied by strong chromospheric heating detectable in the optical bands \citep{2008ApJ...684..644H, VilladsenHallinan2019}. As a result, strong optical flaring does not necessarily imply detectable radio emission within the same observing window.

No causal ordering between optical and radio emission can be inferred from the present data. The radio measurements are averaged over full observing windows and therefore lack the temporal resolution required to establish lead--lag relationships. Where sufficiently high-cadence optical and radio data are available, previous studies have shown that radio emission may precede, follow, or occur independently of optical flares, with delays ranging from seconds to minutes and strong event-to-event variability \citep[e.g.][]{OstenBastian2006, VilladsenHallinan2019}. The absence of any clear ordering in Figure~\ref{fig:radio_optical_lum_window} is therefore consistent with both the limitations of the data and the diversity of behaviours reported in the literature.

The present analysis is insensitive to sub-integration variability and
narrow-band spectral structure owing to the use of time-averaged
UHF SDP continuum imaging. Dynamic spectra and circular polarisation (Stokes~V) measurements would provide key diagnostics of coherent burst emission and may reveal short-duration radio flares that are diluted in the continuum data. While full-Stokes polarimetry in the MeerKAT UHF band is becoming feasible with dedicated calibration approaches (e.g. \textsc{Polkat}; \citealt{2025ascl.soft02026H}), such analysis is not part of the standard SDP continuum products used here and would require dedicated
visibility-level reprocessing beyond the scope of this work. Short-timescale re-imaging of the brightest epochs is currently in preparation (Sandenbergh et~al., in prep.), and future dynamic-spectral and full Stokes analyses will be essential for fully constraining the emission mechanism of LP~888$-$63.

MeerKAT surveys have recently uncovered a growing population of stellar radio transients at sub-mJy to mJy levels, including the serendipitous M-dwarf SCR~1746$-$3214 \citep{2022MNRAS.513.3482A}, the chromospherically active K-subgiant system MKT~J170456.2$-$482100 \citep{2020MNRAS.491..560D}, the X-ray flaring star EXO~040830$-$7134.7 \citep{Driessen2022}, and a recently detected stellar radio flare from a candidate RS~CVn system identified using the TRON pipeline \citep{2025MNRAS.538L..89S}. LP~888$-$63 extends this sample to a mid-M dwarf in a binary environment containing white dwarfs. For context, Table~\ref{tab:comparison} lists representative radio-active stars and previous MeerKAT stellar discoveries alongside LP~888$-$63, illustrating how this source compares in terms of distance, radio luminosity, and observed behaviour.

\begin{table*}
\centering
\caption{Representative radio–active stars and MeerKAT stellar discoveries, for context with LP~888--63. Bands and behaviours are indicative and chosen to highlight emission mechanisms.}
\label{tab:comparison}

\setlength{\tabcolsep}{6.5pt}
\renewcommand{\arraystretch}{1.12}

\begin{tabular*}{\textwidth}{@{} l c c c c p{6.7cm} @{}}
\hline
Object & SpT & Dist. (pc) & Band(s) & $L_{\mathrm{R}}$ (erg\,s$^{-1}$\,Hz$^{-1}$) & Representative radio behaviour \\
\hline
Proxima~Cen & M5.5\,V & 1.30 & mm--cm & $10^{13}$--$10^{14}$ &
Optical and millimetre flares; intermittent cm emission; no persistent highly polarised ECMI bursts (\citealt{MacGregor2018}). \\

UV~Ceti (BL~Ceti) & M5.5\,V & 2.7 & 1--100\,GHz & $10^{13}$--$10^{15}$ &
Mixed incoherent (gyro-) and coherent (ECM) emission; highly polarised decimetric bursts (\citealt{Zic2019, Bastian2022, Plant2024}). \\

AD~Leo & M3.5\,V & 4.9 & 1--2\,GHz & $10^{14}$--$10^{15}$ &
Ultra-bright ($\sim$0.1--0.5\,Jy) coherent bursts; nearly 100\% circularly polarised (\citealt{OstenBastian2006, Zhang2023, Zhang2024b}). \\

SCR~1746$-$3214 & M4\,V & $\sim$12 & L~band & $10^{13}$--$10^{14}$ &
Serendipitous MeerKAT flaring consistent with gyrosynchrotron emission (\citealt{2022MNRAS.513.3482A}). \\

MKT~J170456.2$-$482100 & K\,IV & $\sim$550 & L~band & $10^{13}$--$10^{14}$ &
First MeerKAT transient; sub-mJy variability attributed to stellar/binary origin (\citealt{2020MNRAS.491..560D}). \\

LP~888$-$63 (this work) & M3.5\,V & 23 & UHF (0.816\,GHz) & $(0.2$--$6.6)\times10^{14}$ &
13/41 detections; peak $S_\nu=1.04\pm0.04$\,mJy; $T_{\mathrm{b}}\sim4\times10^{10}$\,K; consistent with gyrosynchrotron (this work). \\
\hline
\end{tabular*}

\end{table*}

Frequent radio flaring in mid–M dwarfs implies efficient particle acceleration and time-variable coronal magnetic fields. This has direct implications for any close-in planet, where enhanced particle and XUV fluxes can erode atmospheres and modulate space weather conditions \citep[e.g.][]{Airapetian2020, Tilley2018,MacGregor2018,Burton2025}. For LP~888$-$63, the next decisive steps are: (i) full-Stokes, multi-band radio monitoring to separate coherent from incoherent components via circular polarisation and spectral indices; (ii) coordinated radio–X-ray–optical campaigns to test placement relative to the Güdel–Benz locus and to search for radio–optical flare simultaneity; and (iii) high-resolution NIR spectroscopy (e.g. FeH) to constrain $v\sin i$ independently of chromospheric lines \citep[e.g.][]{ReinersBasri2006,Reiners2007,Vedantham2022}. Short-timescale analyses with dynamic spectra are particularly powerful in separating coherent and incoherent processes in the MeerKAT band \citep[e.g.][]{Bastian2022,Plant2024}.

Overall, LP~888$-$63 exhibits radio and optical behaviour typical of magnetically active mid–M dwarfs rather than anomalous outliers. Its association with a white–dwarf binary adds environmental interest, but the simplest interpretation, a gyrosynchrotron-dominated stellar corona with possible coherent bursts, remains fully consistent with recent stellar radio results.

\section{Conclusion} \label{con}
We have presented the discovery and characterisation of the radio-flaring M~dwarf LP~888--63 (MKT~J032848.4--271904.6) detected in the LADUMA field using MeerKAT UHF-band observations. The source shows thirteen detections across forty-one epochs with a peak flux density of $1.041 \pm 0.043$~mJy and a brightness temperature of $\sim4\times10^{10}$~K, consistent with gyrosynchrotron emission from mildly relativistic electrons. Proper-motion propagation confirms that the radio source is coincident with the nearby M3.5\,V star LP~888--63 at a distance of $23$~pc, rather than with the adjacent white dwarf LAWD\,14. Optical photometry from \textit{TESS} and MeerLICHT reveals classical white-light flares, although no unambiguous coincident radio--optical flare pair was identified in which an impulsive optical brightening and elevated radio emission were simultaneously confirmed above their respective detection thresholds. The radio measurements represent time-averages over full observing windows of several hours, precluding sub-window timing comparisons; consequently, the presence of short-duration coincident flares cannot be excluded. LP~888--63 thus joins the growing sample of radio-active M~dwarfs uncovered by MeerKAT, demonstrating the power of commensal transient searches within LADUMA for studying magnetic activity in nearby low-mass stars. Future dynamic-spectral and full-Stokes analyses of the LADUMA data, including the short-timescale re-imaging study currently in preparation (Sandenbergh et~al., in prep.), will provide stronger constraints on the emission mechanism and on the temporal relationship between optical and radio flaring in LP~888$-$63.

\section*{Acknowledgements}

The authors thank the ThunderKAT and MeerLICHT teams for useful discussions and support. We are particularly grateful to Gavin Ramsay, Jan van Roestel and Simone Scaringi for their guidance, discussions, and contributions to this work.

This work is based on observations obtained with the MeerKAT radio telescope, a facility of the South African Radio Astronomy Observatory (SARAO), which is a national facility managed by the National Research Foundation (NRF) of South Africa. 
The LADUMA survey forms part of the MeerKAT Large Survey Projects, and we acknowledge the use of data products from the SARAO Science Data Processor (SDP) and the Transient Pipeline (TraP). 
Financial assistance from SARAO towards this research is hereby acknowledged. Optical observations were obtained with MeerLICHT, a collaboration between Radboud University, the University of Cape Town, the University of Oxford, the Netherlands Research School for Astronomy (NOVA), and the South African Astronomical Observatory.
We thank the MeerLICHT operations and data-reduction teams for making these observations available. This work made use of the Inter-University Institute for Data Intensive Astronomy (IDIA) data-intensive research cloud for data processing. IDIA is a South African university partnership involving the University of Cape Town, the University of Pretoria, and the University of the Western Cape.

This work has made use of the SIMBAD and VizieR services provided by the Centre de Données astronomiques de Strasbourg (CDS), Strasbourg, France \citep{2000A&AS..143....9W,2000A&AS..143...23O}. 
We also utilised data from the European Space Agency (\textit{ESA}) mission \textit{Gaia}, processed by the \textit{Gaia} Data Processing and Analysis Consortium (DPAC) \citep{2020yCat.1350....0G,2021A&A...650C...3G}. 
Analysis made use of the \texttt{Astropy} package, a community-developed core Python library for astronomy \citep{2013A&A...558A..33A,2018AJ....156..123A}.

This work is also based on observations collected at the European Southern Observatory and obtained from the ESO science archive.

M.M. acknowledges financial support from the National Research Foundation of South Africa and the National Astrophysics and Space Science Programme (NASSP), which funded this work. PJG is partially supported by NRF SARChI grant 111692.

\section*{Data availability}

The MeerKAT UHF-band data used in this work were obtained from the LADUMA survey and processed using the SARAO Science Data Processor (SDP). Access to the raw visibility data is subject to the LADUMA data access policy and can be requested from the LADUMA collaboration.
The radio image products and transient catalogues generated using the Transient Pipeline (TraP) are available from the corresponding author upon reasonable request.
Optical photometric data from MeerLICHT are available via the MeerLICHT data archive subject to collaboration policies. \textit{TESS} photometry is publicly available from the Mikulski Archive for Space Telescopes (MAST)\footnote{\url{https://mast.stsci.edu}}. Archival ESO spectra used in this work are available through the ESO Science Archive Facility.

Derived data products supporting the findings of this study are available from the corresponding author upon reasonable request.

\bibliographystyle{mnras}
\bibliography{references} 

\appendix

\subsection*{Additional data} \label{Append}

For completeness and reproducibility, we provide in Appendix~\ref{Append} the full observation log for the LADUMA UHF SDPcal-produced epochs used in this work. The table lists the image identifiers and UTC timing information for each epoch, together with the midpoint MJD, the measured RMS noise, and a qualitative data-quality flag used to exclude problematic images from subsequent analysis.

\begin{table*}
\caption{Log of LADUMA UHF SDPcal-produced epochs. The table lists each epoch number, the corresponding image ID, date (UTC), start and end times of the observation (UTC), the Modified Julian Date (MJD) of the midpoint of each observation, the root-mean-square (RMS) noise in mJy\,beam$^{-1}$, and the data quality status.}
\label{tab:observation-log}
\centering
\begin{tabular*}{\textwidth}{@{\extracolsep{\fill}} c c c c c c c c }
\hline
Epoch & Image ID   & Date       & UTC Start Time & UTC End Time & MJD        & RMS        & Status       \\
      &            &            &                &              & (Midpoint) & (mJy/beam) &              \\
\hline
1     & 1637602051 & 2021-11-22 & 17:29:13       & 00:44:22     & 59540.88   & 0.013      & Good         \\
2     & 1640194279 & 2021-12-22 & 17:33:54       & 00:14:58     & 59570.88   & 0.017      & Bad          \\
3     & 1640885597 & 2021-12-30 & 17:34:59       & 23:41:59     & 59579.37   & 0.018      & Bad          \\
4     & 1640972173 & 2021-12-31 & 17:38:09       & 23:45:09     & 59580.37   & 0.015      & Good         \\
5     & 1641490520 & 2022-01-06 & 17:38:12       & 23:44:24     & 59586.37   & 0.014      & Good         \\
6     & 1642093878 & 2022-01-13 & 17:13:09       & 23:22:38     & 59593.35   & 0.014      & Good         \\
7     & 1642181477 & 2022-01-14 & 17:33:12       & 23:05:27     & 59594.35   & 0.014      & Good         \\
8     & 1642267950 & 2022-01-15 & 17:34:21       & 23:06:21     & 59595.35   & 0.015      & Good         \\
9     & 1642613170 & 2022-01-19 & 17:27:42       & 22:25:36     & 59599.34   & 0.014      & Good         \\
10    & 1642700870 & 2022-01-20 & 17:48:15       & 22:40:54     & 59600.35   & 0.015      & Good         \\
11    & 1643715069 & 2022-02-01 & 11:32:45       & 21:31:28     & 59612.19   & ---        & Visually Bad \\
12    & 1644059590 & 2022-02-05 & 11:14:50       & 21:15:15     & 59616.18   & 0.013      & Good         \\
13    & 1644318129 & 2022-02-08 & 11:03:53       & 21:04:02     & 59619.17   & 0.014      & Good         \\
14    & 1644752473 & 2022-02-13 & 11:42:54       & 21:38:51     & 59624.20   & 0.013      & Good         \\
15    & 1644576673 & 2022-02-11 & 10:52:46       & 20:53:04     & 59622.17   & 0.013      & Good         \\
16    & 1658030772 & 2022-07-17 & 04:04:49       & 10:35:12     & 59777.81   & 0.013      & Good         \\
17    & 1658192293 & 2022-07-19 & 00:59:52       & 08:37:23     & 59779.71   & 0.013      & Good         \\
18    & 1658449093 & 2022-07-22 & 00:19:29       & 06:03:43     & 59782.64   & 0.016      & Good         \\
19    & 1658966171 & 2022-07-27 & 23:56:44       & 07:28:12     & 59788.16   & 0.014      & Good         \\
20    & 1661046814 & 2022-08-21 & 01:54:56       & 09:27:11     & 59812.74   & 0.015      & Good         \\
21    & 1661391366 & 2022-08-25 & 01:37:31       & 08:42:48     & 59816.72   & 0.582      & Visually Bad \\
22    & 1661637077 & 2022-08-27 & 21:52:37       & 07:05:40     & 59819.11   & 0.012      & Good         \\
23    & 1661895557 & 2022-08-30 & 21:40:38       & 05:12:37     & 59822.07   & 0.013      & Good         \\
24    & 1662326779 & 2022-09-04 & 21:27:39       & 06:57:33     & 59827.10   & 0.012      & Good         \\
25    & 1662843383 & 2022-09-10 & 20:57:37       & 06:28:50     & 59833.08   & 0.014      & Good         \\
26    & 1662767184 & 2022-09-09 & 23:47:47       & 07:26:21     & 59832.16   & 0.012      & Good         \\
27    & 1663015699 & 2022-09-12 & 20:49:42       & 06:20:39     & 59835.07   & 0.012      & Good         \\
28    & 1663188078 & 2022-09-14 & 20:42:38       & 06:13:43     & 59837.07   & 0.013      & Good         \\
29    & 1663274258 & 2022-09-15 & 20:38:57       & 06:08:50     & 59838.06   & 0.013      & Good         \\
30    & 1663368199 & 2022-09-16 & 22:44:38       & 07:12:33     & 59839.13   & 0.013      & Good         \\
31    & 1663619479 & 2022-09-19 & 20:32:44       & 06:04:04     & 59842.06   & 0.013      & Good         \\
32    & 1663791375 & 2022-09-21 & 20:17:36       & 05:48:41     & 59844.05   & 0.043      & Bad          \\
33    & 1663963575 & 2022-09-23 & 20:07:39       & 05:38:44     & 59846.04   & 0.015      & Good         \\
34    & 1664049680 & 2022-09-24 & 20:02:39       & 02:10:42     & 59846.97   & 0.017      & Bad          \\
35    & 1664228310 & 2022-09-26 & 21:39:48       & 07:11:17     & 59849.11   & 0.013      & Good         \\
36    & 1664480802 & 2022-09-29 & 19:48:04       & 05:17:58     & 59852.03   & 0.014      & Good         \\
37    & 1664653281 & 2022-10-01 & 19:42:41       & 05:12:34     & 59854.02   & 0.013      & Good         \\
38    & 1664739015 & 2022-10-02 & 19:31:33       & 05:01:27     & 59855.02   & 0.012      & Good         \\
39    & 1665175572 & 2022-10-07 & 20:47:41       & 06:20:05     & 59860.07   & 0.012      & Good         \\
40    & 1665255975 & 2022-10-08 & 19:07:39       & 04:37:25     & 59861.00   & 0.012      & Good         \\
41    & 1665688881 & 2022-10-13 & 19:22:43       & 04:58:24     & 59866.01   & 0.013      & Good         \\
42    & 1665864898 & 2022-10-15 & 20:16:16       & 03:53:16     & 59868.01   & 0.013      & Good         \\
43    & 1665948077 & 2022-10-16 & 19:22:32       & 04:58:13     & 59869.01   & 0.012      & Good         \\
44    & 1666553482 & 2022-10-23 & 19:32:47       & 05:04:39     & 59876.02   & 0.027      & Bad          \\
45    & 1667331070 & 2022-11-01 & 19:32:27       & 05:01:10     & 59885.02   & 0.013      & Good         \\
46    & 1667675546 & 2022-11-05 & 19:13:44       & 04:41:08     & 59889.00   & 0.013      & Good         \\
47    & 1667849352 & 2022-11-07 & 19:30:31       & 04:28:19     & 59891.00   & 0.012      & Good         \\
48    & 1668271473 & 2022-11-12 & 16:45:58       & 01:43:53     & 59895.89   & 0.014      & Good         \\
\hline
\end{tabular*}
\end{table*}

\section{Supplementary optical and radio analysis} 

This appendix presents supplementary figures that support the window-based optical–radio analysis described in Section \ref{MeelL}. These figures provide additional visual context for the construction of the window-averaged luminosity measurements and illustrate the diversity of optical and radio behaviour observed on epoch timescales.

\subsection{Window-level optical and radio light curves}

Figures~A1--A6 show zoomed MeerLICHT multi-band optical light curves for all MeerKAT observing windows that contain contemporaneous MeerLICHT observations, together with the corresponding radio measurements. In each case, the MeerKAT radio flux density represents a time-averaged measurement over the full duration of the observing window and is plotted at the temporal midpoint. Optical light curves are shown relative to the quiescent baseline, allowing flare activity above quiescence to be readily identified.

The six windows illustrate a range of behaviours, including radio detections accompanied by pronounced multi-band optical flaring, radio detections occurring during optically quiescent intervals, and optical variability present during radio non-detections.

\begin{figure}
    \centering
    \includegraphics[width=\linewidth]{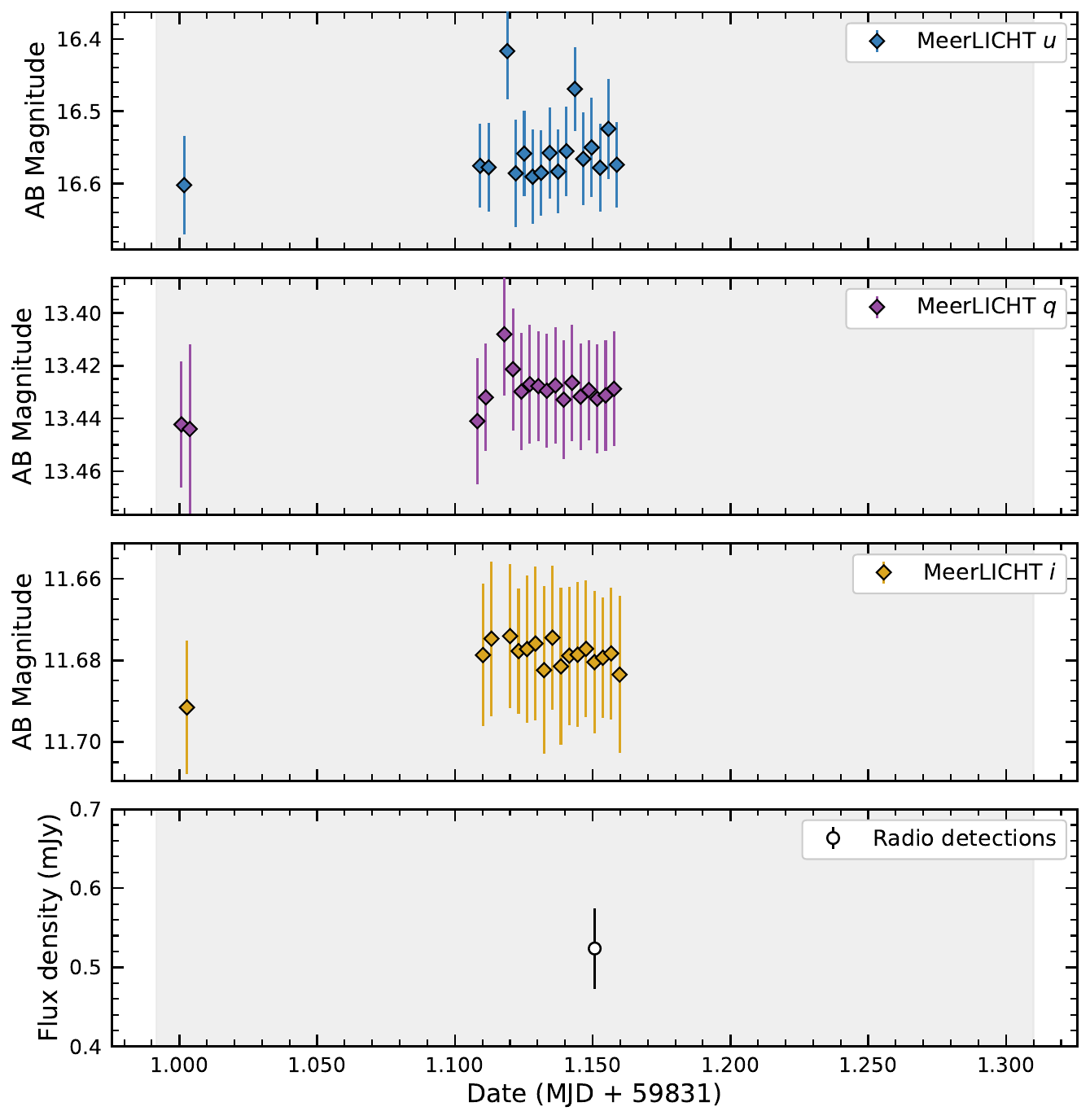}
    \caption{Window~24 (59831.99152--59832.30997~MJD): zoomed MeerLICHT multi-band optical light curves relative to the quiescent baseline, together with the corresponding time-averaged MeerKAT radio measurement.}
    \label{fig:app_window_24}
\end{figure}

\begin{figure}
    \centering
    \includegraphics[width=\linewidth]{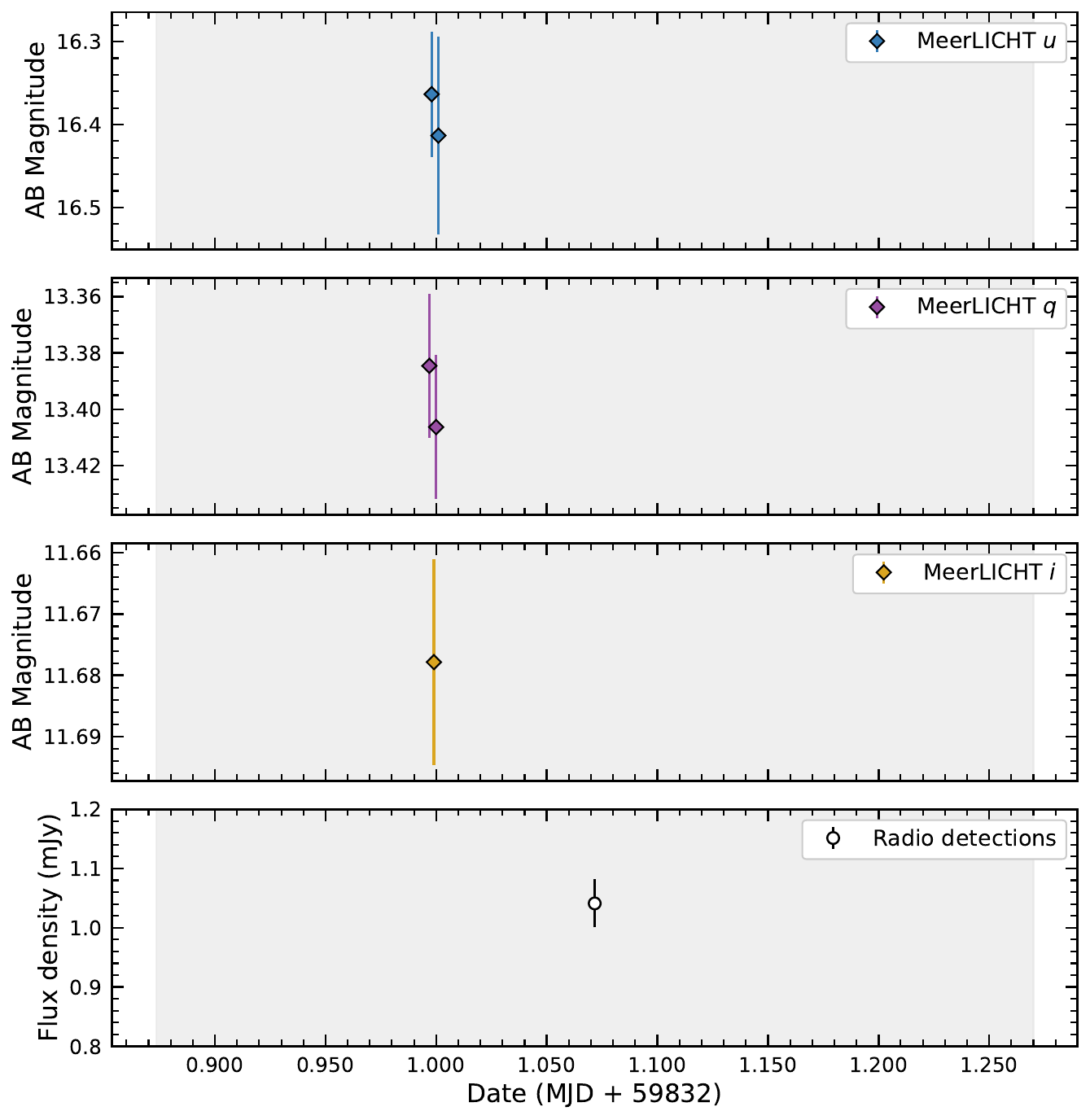}
    \caption{As in Figure~\ref{fig:app_window_24}, but for Window~23 (59832.87334--59833.27002~MJD).}
    \label{fig:app_window_23}
\end{figure}

\begin{figure}
    \centering
    \includegraphics[width=\linewidth]{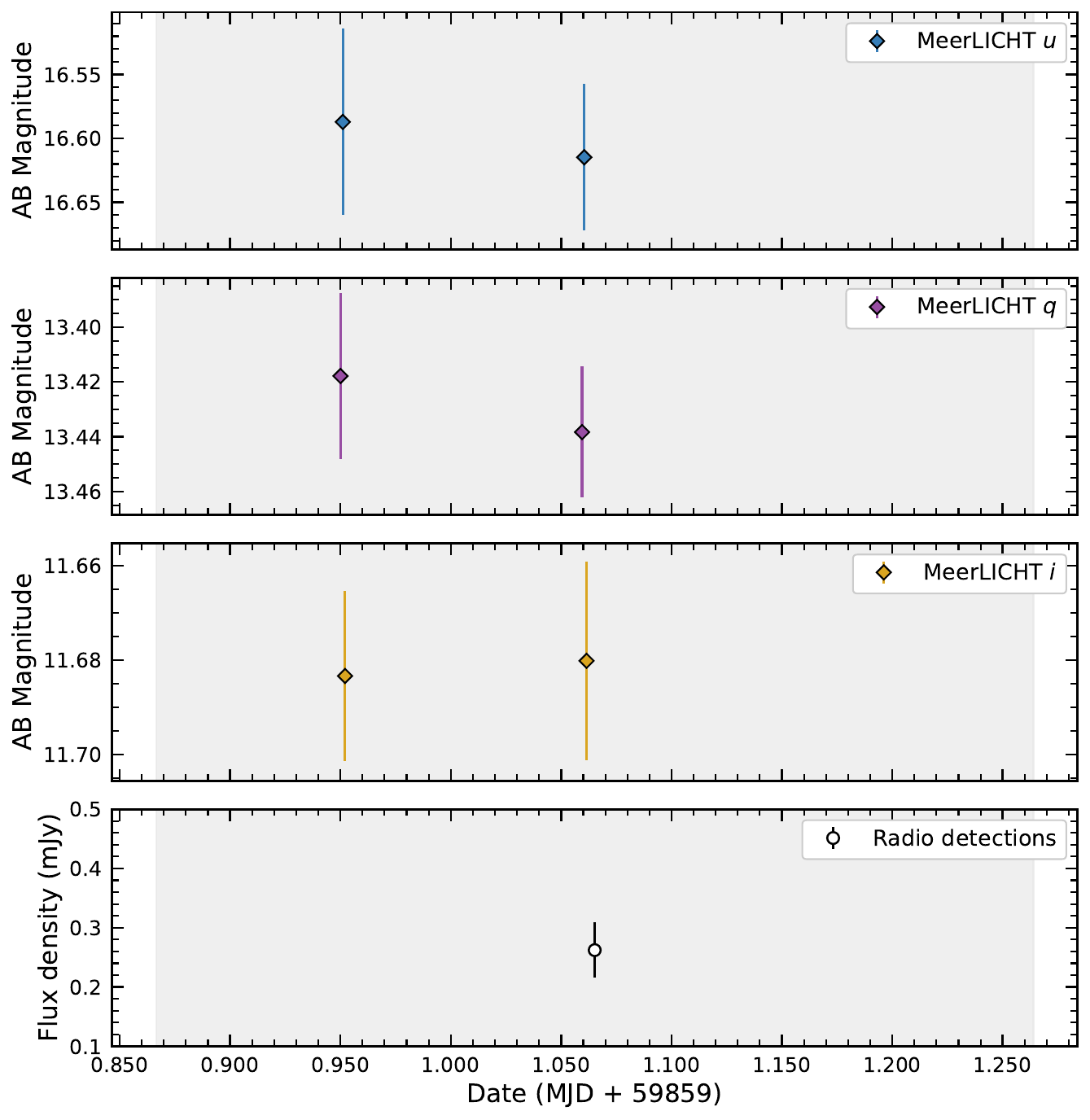}
    \caption{As in Figure~\ref{fig:app_window_24}, but for Window~35 (59859.86645--59860.26395~MJD).}
    \label{fig:app_window_35}
\end{figure}

\begin{figure}
    \centering
    \includegraphics[width=\linewidth]{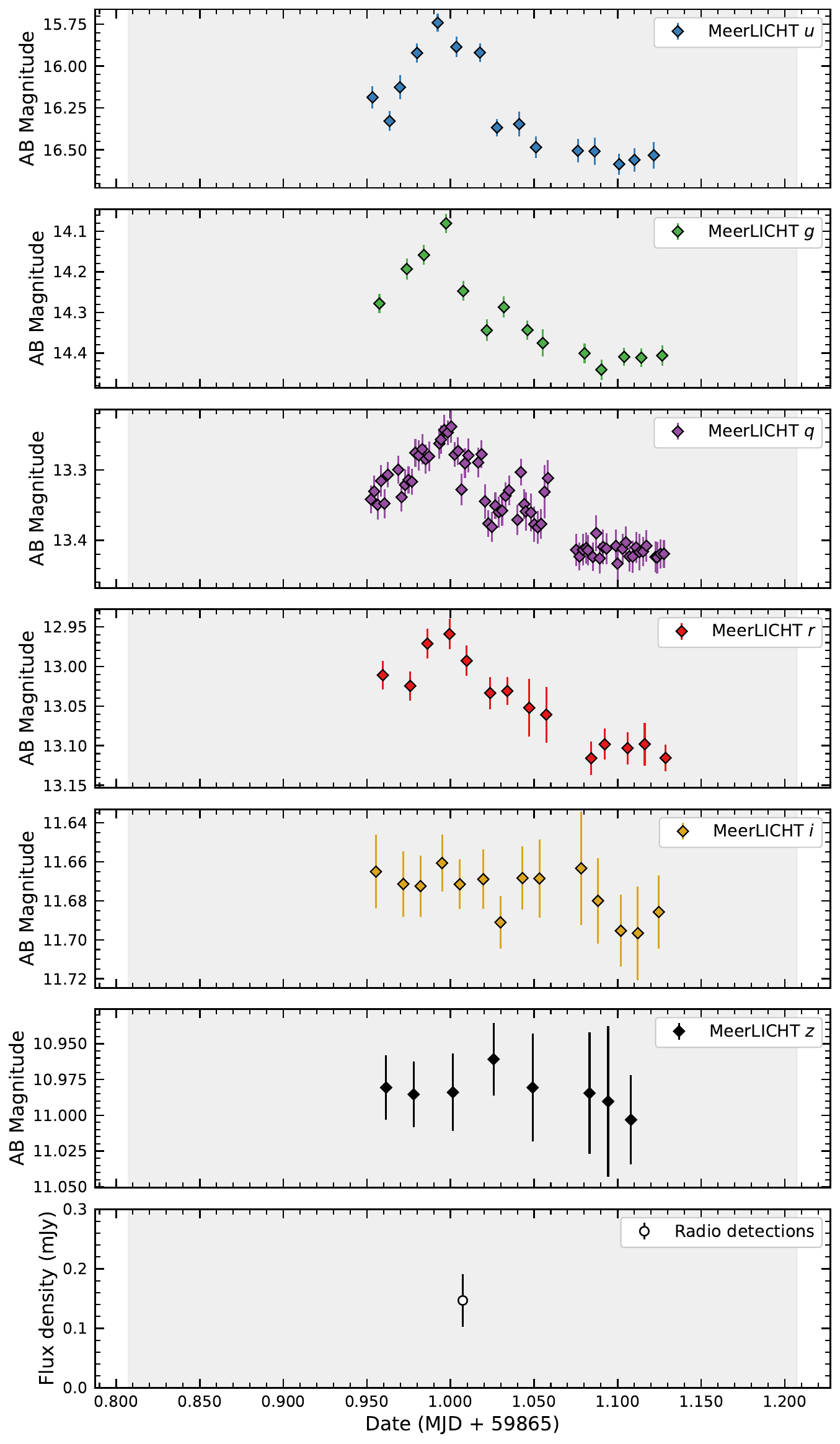}
    \caption{As in Figure~\ref{fig:app_window_24}, but for Window~37 (59865.80744--59866.20722~MJD).}
    \label{fig:app_window_37}
\end{figure}

\begin{figure}
    \centering
    \includegraphics[width=\linewidth]{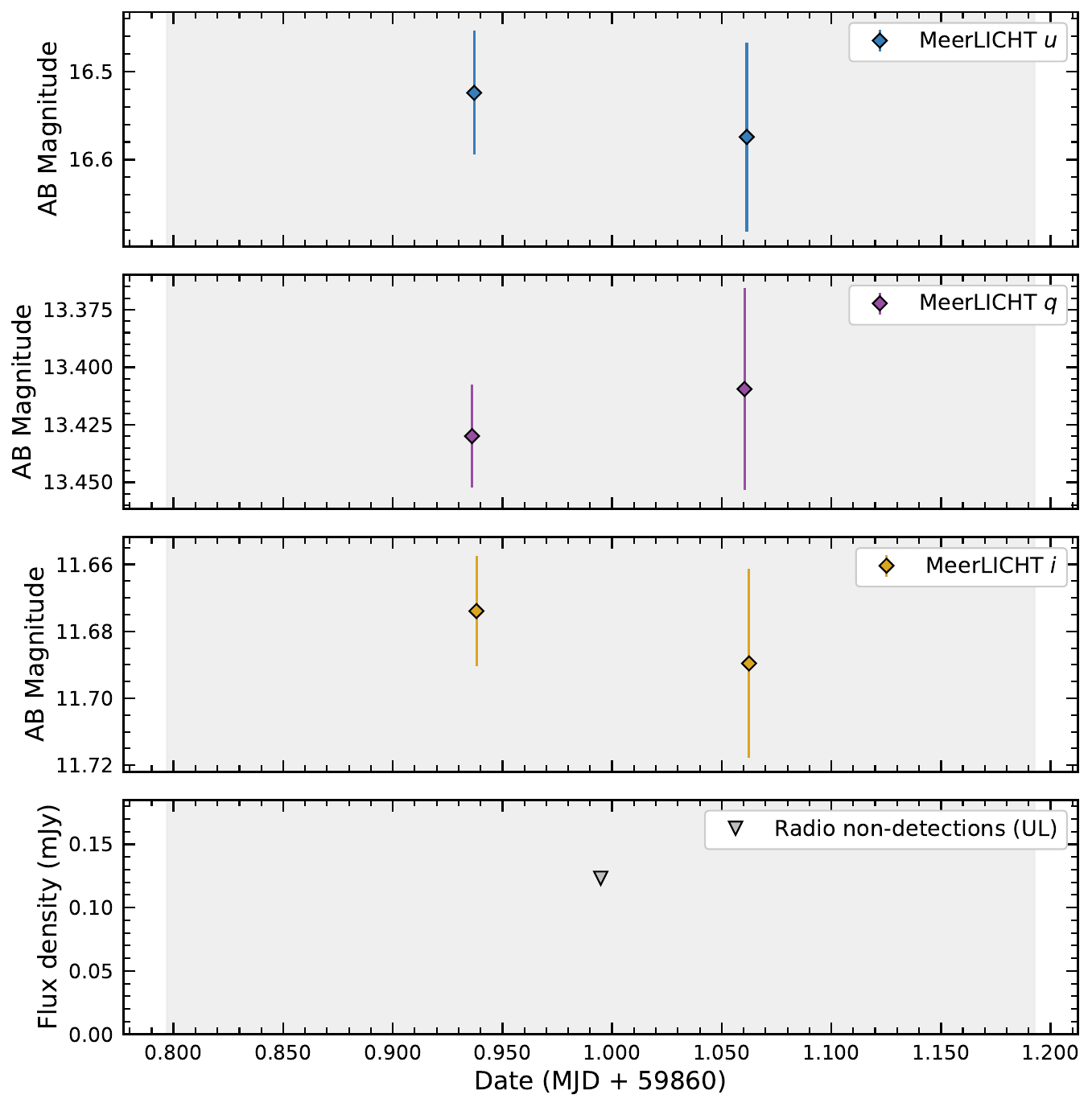}
    \caption{As in Figure~\ref{fig:app_window_24}, but for Window~36 (59860.79698--59861.19265~MJD), which contains a radio non-detection.}
    \label{fig:app_window_36}
\end{figure}

\begin{figure}
    \centering
    \includegraphics[width=\linewidth]{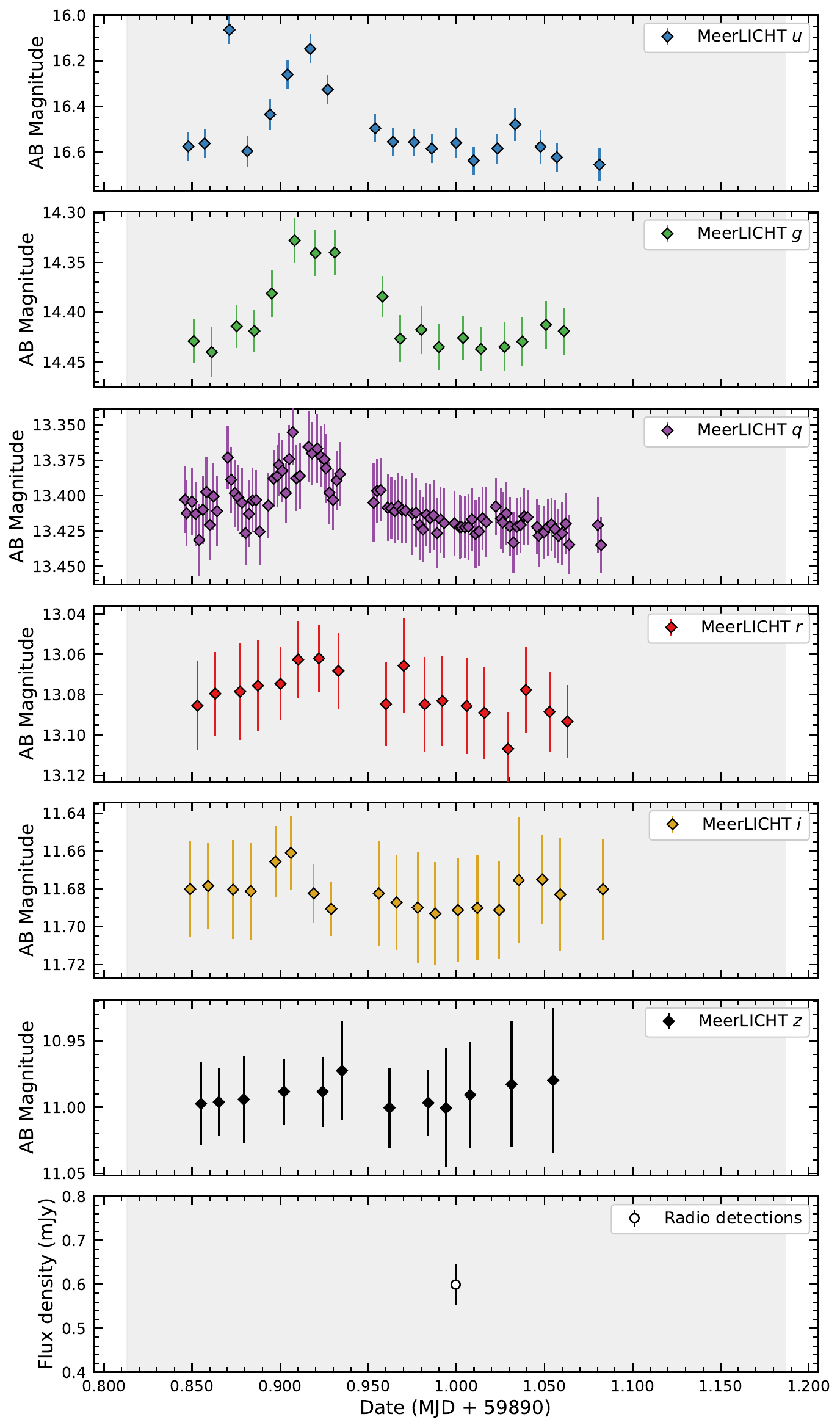}
    \caption{As in Figure~\ref{fig:app_window_24}, but for Window~42 (59890.81286--59891.18633~MJD).}
    \label{fig:app_window_42}
\end{figure}

\subsection{Peak optical flare luminosity comparison}

\begin{figure}
    \centering
    \includegraphics[width=\linewidth]{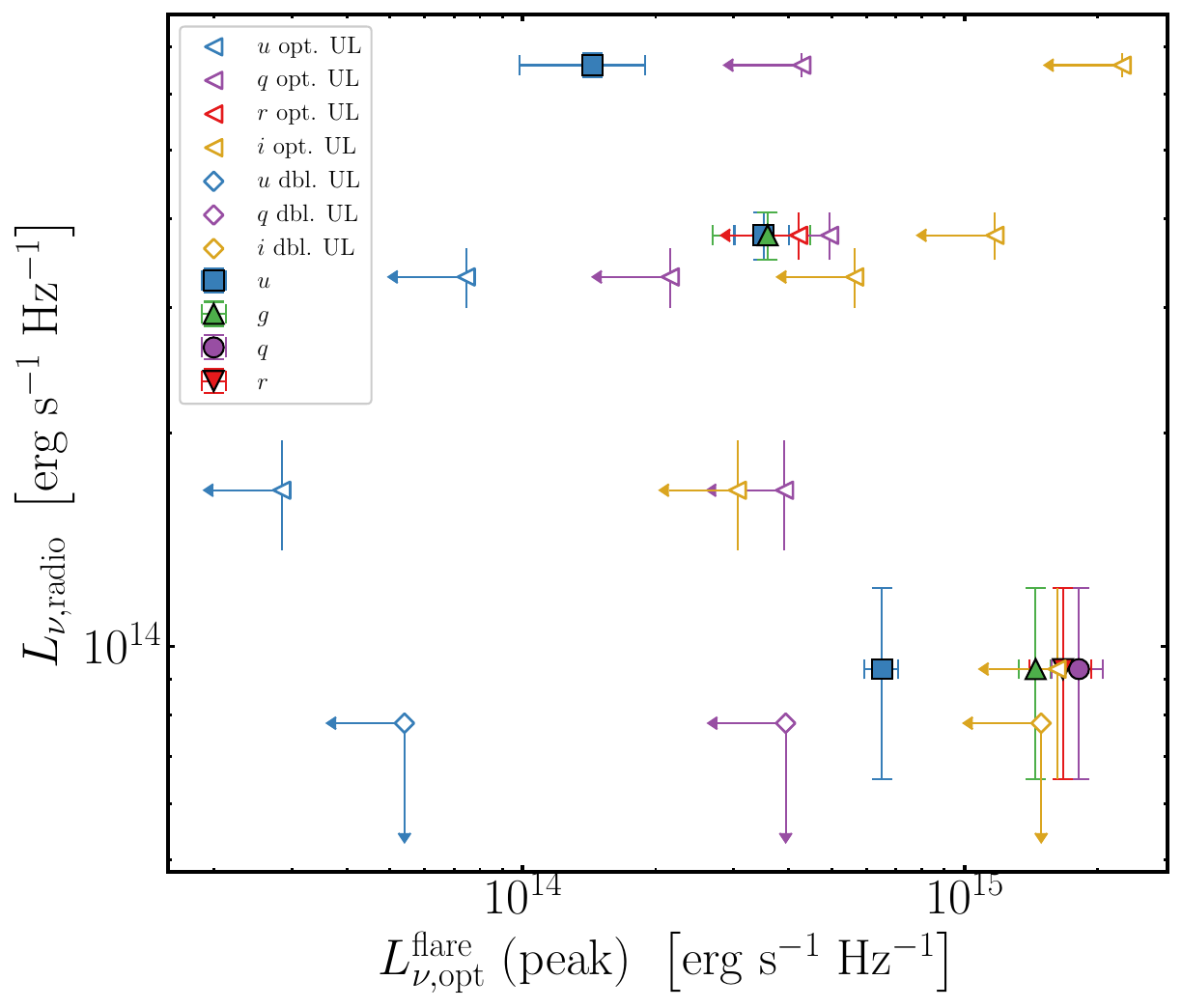}
    \caption{As in Figure \ref{fig:radio_optical_lum_window}, but showing the peak (rather than window-averaged) optical flare luminosity for windows with a significant optical flare excess. Since an optically quiescent window has no defined flare peak, windows without a significant excess are shown at the same 3$\sigma$ optical flare-luminosity upper limit used in Figure \ref{fig:radio_optical_lum_window}. Marker and colour conventions follow Figure 9. Error bars represent measurement uncertainties in both luminosities.}
    \label{fig:app_peak_lum}
\end{figure}

Figure \ref{fig:app_peak_lum} shows the peak optical flare luminosity measured within each overlapping observing window as a function of the corresponding window-averaged radio luminosity. This comparison complements the window-averaged luminosity analysis presented in the main text by highlighting the most optically energetic flares occurring during each epoch. As discussed in Section \ref{MeelL}, this diagnostic compares a peak optical quantity with a time-averaged radio measurement and is therefore included as supplementary information.

\end{document}